\documentclass[preprint2]{aastex}

\slugcomment{}

\shorttitle{Arc and jet clouds toward HESS J1023$-$575}
\shortauthors{Furukawa et al.}

\begin{document}

\title{The jet and arc molecular clouds toward Westerlund 2, RCW 49, and HESS J1023$-$575; $^{12}$CO and $^{13}$CO ($J$=2--1 and $J$=1--0) observations with NANTEN2 and Mopra Telescope}


\author{
N.~Furukawa\altaffilmark{1, 2},
A.~Ohama\altaffilmark{1},
T.~Fukuda\altaffilmark{1},
K.~Torii\altaffilmark{1},
T.~Hayakawa\altaffilmark{1},
H.~Sano\altaffilmark{1},
T.~Okuda\altaffilmark{1, 3},
H.~Yamamoto\altaffilmark{1},
N.~Moribe\altaffilmark{1}, 
A.~Mizuno\altaffilmark{4}, 
H.~Maezawa\altaffilmark{5}, 
T.~Onishi\altaffilmark{5}, 
A.~Kawamura\altaffilmark{6}, 
N.~Mizuno\altaffilmark{6}, 
J.R.~Dawson\altaffilmark{7, 8}, 
T.M.~Dame\altaffilmark{9}, 
Y.~Yonekura\altaffilmark{10},
F.~Aharonian\altaffilmark{11, 12}, 
E.~de~O\~{n}a~Wilhelmi\altaffilmark{12}, 
G.P.~Rowell\altaffilmark{13},
R.~Matsumoto\altaffilmark{14}, 
Y.~Asahina\altaffilmark{14},
and
Y.~Fukui\altaffilmark{1}
}

\altaffiltext{1}{Department of Physics, Nagoya University, Furo-cho, Chikusa-ku, Nagoya 464-8602, Japan; naoko@a.phys.nagoya-u.ac.jp, fukui@a.phys.nagoya-u.ac.jp}
\altaffiltext{2}{present address: Center for Astronomy, Ibaraki University, 2-1-1 Bunkyo, Mito, Ibaraki 310-8512, Japan}
\altaffiltext{3}{present address: National Astronomical Observatory of Japan, Osawa, Mitaka, Tokyo 181-8588, Japan}
\altaffiltext{4}{Solar-Terrestrial Environment Laboratory, Nagoya University, Furo-cho, Chikusa-ku, Nagoya 464-8601, Japan}
\altaffiltext{5}{Department of Astrophysics, Graduate School of Science, Osaka Prefecture University, Sakai, Osaka 599-8531, Japan}
\altaffiltext{6}{National Astronomical Observatory of Japan, Osawa, Mitaka, Tokyo 181-8588, Japan}
\altaffiltext{7}{School of Mathematics and Physics, University of Tasmania, Sandy Bay Campus, Churchill Avenue, Sandy Bay, TAS 7005, Australia}
\altaffiltext{8}{present address: Australia Telescope National Facility, CSIRO Astronomy and Space Science, PO Box 76, Epping, NSW 1710, Australia}
\altaffiltext{9}{Harvard-Smithsonian Center for Astrophysics, 60 Garden Street, Cambridge, MA 02138, USA}
\altaffiltext{10}{Center for Astronomy, Ibaraki University, 2-1-1 Bunkyo, Mito, Ibaraki 310-8512}
\altaffiltext{11}{Dublin Institute for Advanced Studies, 31 Fitzwiliam Place, Dublin 2, Ireland}
\altaffiltext{12}{Max-Planck-Institut f\"ur Kernphysik, P.O. Box 103980, D 69029 Heidelberg, Germany}
\altaffiltext{13}{School of Chemistry \& Physics, University of Adelaide, Adelaide 5005, Australia}
\altaffiltext{14}{Department of Physics, Graduate School of Science, Chiba University, 1-33 Yayoi-cho, Inage-ku, Chiba 263-8522, Japan}

\begin{abstract}

We have made new CO observations of two molecular clouds, which we call ``{\it jet}'' and ``{\it arc}'' clouds, toward the stellar cluster Westerlund 2 and the TeV $\gamma$-ray source HESS J1023$-$575.
The {\it jet} cloud shows a linear structure from the position of Westerlund 2 on the east.
In addition, we have found a new counter {\it jet} cloud on the west.
The {\it arc} cloud shows a crescent shape in the west of HESS J1023$-$575.
A sign of star formation is found at the edge of the {\it jet} cloud and gives a constraint on the age of the {\it jet} cloud to be $\sim$Myrs.
An analysis with the multi CO transitions gives temperature as high as 20\,K in a few places of the {\it jet} cloud, suggesting that some additional heating may be operating locally.
The new TeV $\gamma$-ray images by H.E.S.S.\ correspond to the {\it jet} and {\it arc} clouds spatially better than the giant molecular clouds associated with Westerlund 2.
We suggest that the {\it jet} and {\it arc} clouds are not physically linked with Westerlund 2 but are located at a greater distance around 7.5\,kpc.
A microquasar with long-term activity may be able to offer a possible engine to form the {\it jet} and {\it arc} clouds and to produce the TeV $\gamma$-rays, although none of the known microquasars have a Myr age or steady TeV $\gamma$-rays.
Alternatively, an anisotropic supernova explosion which occurred $\sim$Myr ago may be able to form the {\it jet} and {\it arc} clouds, whereas the TeV $\gamma$-ray emission requires a microquasar formed after the explosion.

\end{abstract}

\keywords{ISM: clouds --- ISM: individual objects: jet and arc molecular clouds, HESS J1023$-$575 --- Stars: individual: Westerlund 2 }


\section{INTRODUCTION}\label{sec:introduction}
High-mass stars influence the interstellar medium (ISM) throughout their lifetime both in physical and chemical contexts.
High-mass stars compress the surrounding ISM via \ion{H}{2} regions, stellar winds, and supernova (SN) explosions and may trigger subsequent star formation \citep[see e.g.,][]{1999PASJ...51..791Y, 2006A&A...446..171Z, 2010A&A...523A...6D, 2011ApJ...728..127D}.
The SN explosions also inject heavy elements into the interstellar space and form compact stellar remnants like neutron stars and black-holes, leading to high-energy objects such as microquasars and pulsar wind nebulae (PWNe).
Including all these processes, high-mass stars play a crucial role in the galactic evolution through the tight linkage with various energetic events. In spite of the importance of the high-mass stars, a number of astrophysical issues yet remain unanswered. One of them is cosmic ray (CR) acceleration.
While supernova remnants (SNRs) (e.g., RX J1713.7$-$3946: \citealt{2006A&A...449..223A, 2007A&A...464..235A}; RX J0852.0$-$4622: \citealt{2007ApJ...661..236A}) and PWNe (e.g., HESS J1825$-$137: \citealt{2006A&A...460..365A}; Vela X: \citealt{2012A&A...548A..38A}) are generally believed to be the most plausible sites of CR acceleration via the diffusive shock acceleration \citep[e.g.,][]{1978MNRAS.182..147B, 1978ApJ...221L..29B}, it is also expected that young stellar clusters \citep[e.g., Westerlund 1:][]{2012A&A...537A.114A}, $\gamma$-ray binaries (e.g., LS5039: \citealt{2006A&A...460..743A}; LS I+61{\degr}303: \citealt{2006Sci...312.1771A}), and microquasars \citep[e.g.,][]{2005A&A...432..609B} may be additional acceleration sites.
The CR electrons emit $\gamma$-rays via inverse Compton scattering with ambient radiation fields (called leptonic origin) and, on the other hand, the CR protons emit $\gamma$-rays by interacting with protons in their surrounding medium (called hadronic origin).
Recent progress of the high resolution $\gamma$-ray observations are revealing new aspects of CR acceleration sites as offered by the $\gamma$-ray telescopes including H.E.S.S.\ \citep{2005Sci...307.1938A}, MAGIC \citep{2004NewAR..48..339L}, VERITAS \citep{2006APh....25..391H}, Fermi \citep{2010RPPh...73g4901M}, and AGILE \citep{2009A&A...502..995T}.

The very high energy $\gamma$-ray survey for the Galactic plane by High Energy Stereoscopic System (H.E.S.S.) unveiled an extended TeV $\gamma$-ray source, HESS J1023$-$575 \citep{2007A&A...467.1075A} toward the young rich stellar cluster Westerlund 2 (hereafter Wd 2).
Wd 2 is one of the rare super star clusters in the Galaxy associated with the \ion{H}{2} region RCW 49 \citep{2010ARA&A..48..431P} and its stellar mass and age are estimated to be $\sim10^4\, M_\sun$ \citep{2007A&A...466..137A, 2007A&A...463..981R} and $\sim$ 2 -- 3\,Myr \citep{1998A&AS..127..423P}.
Since no SNRs and pulsars were known at the time of the discovery, it was discussed that the $\gamma$-ray may be due to the dynamical activity like the stellar-wind collision from the cluster \citep{2007A&A...467.1075A}, and theoretical models of $\gamma$-rays production via the stellar wind were discussed \citep[e.g.,][]{2007MNRAS.382..367B, 2007A&A...474..689M}.
\citet{2007A&A...467.1075A} also discussed an alternative that a blister of the \ion{H}{2} region RCW 49 \citep{1997A&A...317..563W} is a candidate for the $\gamma$-ray origin, since the peak position of the $\gamma$-ray source is shifted by 8 arcmin from the center of Wd 2 and coincident with the blister, noting that the size of the $\gamma$-ray source may be too large to be formed by the stellar-wind collisions from the WRs and O-type stars.
Subsequently, analyses of the {\it Suzaku} X-ray spectra indicate that $\alpha$-elements are over abundant compared with iron, suggesting that the ISM may have experienced some SN explosions \citep{2009PASJ...61.1229F}.
\citet{2009ApJ...707L.179F} argued that the shock front of an SNR occurred in the past may have accelerated CRs and emit the TeV $\gamma$-rays by the interaction with the molecular gas toward Wd 2.
Recently, {\it Fermi} collaboration discovered a $\gamma$-ray pulsar PSR J1023$-$5746 near the peak position of the TeV $\gamma$-ray source \citep{2010ApJ...725..571S}.
\citet{2011ApJ...726...35A} suggested that HESS J1023$-$575 was an associated PWN.
The {\it Suzaku} X-ray observations, however, did not detect pronounced synchrotron emission toward the Wd 2 region, whereas the non-thermal X-rays are a general characteristic of PWNe \citep{2009PASJ...61.1229F}.
The interpretation in terms of the PWN thus remains yet unsettled.
Most recently, \citet{2011A&A...525A..46H} made new H.E.S.S.\ observations toward Wd 2 with 3.5 times more integration time than \citet{2007A&A...467.1075A}.
These new observations have higher sensitivity than the earlier observations and presented the TeV $\gamma$-ray distribution in the two energy bands, 0.7 -- 2.5\,TeV and $> 2.5$\,TeV, as well as a few additional features within $1\arcdeg$ of Wd 2.

In order to search for molecular gas which is associated with Wd 2 and the $\gamma$-ray source, \citet[hereafter Paper I]{2009PASJ...61L..23F} studied molecular clouds by using the $^{12}$CO ($J$=1--0) transition at 2.6 arcmin resolution from the NANTEN galactic plane survey (NGPS) \citep{2004ASPC..317...59M}.
The CO emission in the region is fairly complicated since the region is close to the tangential point of the Carina arm at $l \sim 280\degr$ \citep[e.g.,][]{1987ApJ...315..122G}.
The region has several CO components in a range from $-10$\,km\,s$^{-1}$ to 30\,km\,s$^{-1}$ \citep{2007ApJ...665L.163D, 2009ApJ...696L.115F}.
\citet{2009ApJ...696L.115F} and \citet{2010ApJ...709..975O} showed that two giant molecular clouds (GMCs) at 4\,km\,s$^{-1}$ and 16\,km\,s$^{-1}$ are physically associated with Wd 2 and RCW 49 as supported by the morphological correspondence and physical association with RCW 49, while the more spatially extended GMC peaked at 11\,km\,s$^{-1}$ is not directly associated with Wd 2 on a 10-pc scale.
Paper I reported the discovery of two additional unusual CO clouds in a velocity range 20 -- 30\,km\,s$^{-1}$ over a degree-scale field. These two clouds are named a ``{\it jet}'' cloud and an ``{\it arc}'' cloud and are distinguished with the italic type from general astrophysical jets such as the microquasar jet.
The {\it jet} cloud shows a straight feature across $\sim1\degr$ and is extended only toward the east (in the Galactic coordinates).
The {\it arc} cloud is located in the west of the TeV $\gamma$-ray source and has a crescent shape whose center is apparently located toward the center of the TeV $\gamma$-ray source.
The association among the {\it jet} cloud, the {\it arc} cloud, and HESS J1023$-$575 is thus strongly suggested, whereas the association between the other two GMCs \citep{2009ApJ...696L.115F} and HESS J1023$-$575 is open.
In Paper I the authors hypothesized that a microquasar jet with an SN explosion or an anisotropic SN explosion may have created the {\it jet} and {\it arc} clouds. In the both hypotheses, it was discussed that a spherical explosion formed the {\it arc} cloud, and the high-energy jet phenomenon compressed the \ion{H}{1} gas cylindrically along the {\it jet} cloud axis to form the {\it jet} cloud.
Concerning the cloud formation by the high-energy jet phenomenon, similar molecular clouds are reported toward the microquasar SS 433 and toward an unidentified driving object at $(l , b) =$ ($348\fdg5$, $-2\degr$ -- $+2\degr$) \citep{2008PASJ...60..715Y}.
Most recently, hydrodynamical numerical simulations have been carried out on such a process, demonstrating that the hypothesis offers a possible explanation for the formation of the {\it jet} cloud (Asahina et al.\ 2013, in preparation).

Physical association among all the objects concerned, the {\it jet} and {\it arc} clouds, Wd 2, HESS J1023$-$575, and PSR J1023$-$5746, are still uncertain.
Distance determination of Wd 2 is not settled.
Spectro-photometry of WR 20a and O stars gives a distance of $8.0\pm1.4$\,kpc \citep{2005A&A...432..985R, 2007A&A...463..981R, 2011A&A...535A..40R}, while \mbox{$J\!H\!K_{\rm s}$} observations give 2.8\,kpc \citep{2007A&A...466..137A}.
Alternative distance estimates by using a flat Galactic rotation curve \citep{1993A&A...275...67B} independently yield $6.0\pm1.0$\,kpc based on the GMC at 11\,km\,s$^{-1}$ \citep{2007ApJ...665L.163D}, and $5.4^{+1.1}_{-1.4}$\,kpc based on the two GMCs at 4\,km\,s$^{-1}$ and 16\,km\,s$^{-1}$ identified as the parent GMCs \citep{2009ApJ...696L.115F, 2010ApJ...709..975O}.
We shall adopt 5.4\,kpc as the distance to Wd 2 in this paper.
On the other hand, the central velocity of the {\it jet} and {\it arc} clouds is $\sim$26\,km\,s$^{-1}$.
If we adopt the same Galactic rotation curve, the kinematic distance of the {\it jet} and {\it arc} clouds is 7.5\,kpc which is different from Wd 2 by $\sim$2.1\,kpc.
The {\it jet} and {\it arc} clouds may not be physically related to Wd 2 and the $\gamma$-ray pulsar PSR J1023$-$5746 may not be necessarily connected with Wd 2, either.
If the $\gamma$-ray pulsar is located beyond 6.0\,kpc, the velocity of the pulsar required for escaping from the cluster has to be as large as 3000\,km\,s$^{-1}$ or more for a characteristic age of 4.6\,kyr.
\citet{2010ApJ...725..571S} and \citet{2011ApJ...726...35A} therefore suggested the distance to be 1.8 -- 2.4\,kpc.

The aim of the present work is to better understand detailed distribution and physical properties of the {\it jet} and {\it arc} clouds by observing CO at angular resolutions of 47{\arcsec} -- 100{\arcsec}  higher than that of 160{\arcsec} used in Paper I.
Section \ref{sec:observations} gives observations and Section \ref{sec:results} presents the results.
In the Section \ref{sec:discussion}, we reexamine physical association among all the objects toward the Wd 2 region, the formation scenarios of the {\it jet} and {\it arc} clouds, and the origin of the $\gamma$-ray emission.
We summarize the present work in Section \ref{sec:conclusion}.


\section{OBSERVATIONS}\label{sec:observations}

We list all the observed transitions in Table \ref{prop_tel}.
In the followings we describe details of individual observations with the NANTEN2 4\,m and Mopra 22\,m telescopes.
Observed parameters in each transition except for $^{13}$CO($J$=2--1) are summarized in Tables \ref{obs_param} -- \ref{obs_para13co10}.

\subsection{$^{12}$CO($J$=2--1) and $^{13}$CO($J$=2--1) observations with NANTEN2}
Observations in the $^{12}$CO($J$=2--1) transition (230.538\,GHz) were carried out using the NANTEN2 4\,m telescope at Atacama in Chile in 2008 October.
The beam size (HPBW) was 90{\arcsec} at 230\,GHz corresponding to 3.3\,pc at a distance of 7.5\,kpc.
We used a 4\,K cooled superconductor-insulator-superconductor (SIS) mixer receiver whose typical system temperature at the zenith including the atmosphere was $\sim$300\,K in the single-side-band (SSB). 
The spectrometer was a wide band Acousto-Optical Spectrometer (AOS) with a bandwidth of 250\,MHz (325\,km\,s$^{-1}$ in velocity), a frequency spacing per channel of 145 kHz (0.19\,km\,s$^{-1}$), and an effective frequency resolution of 250\,kHz (0.33\,km\,s$^{-1}$) at 230 GHz.
The spectroscopic data were calibrated to the corrected antenna temperature $T_{\rm A}^{\ast}$ by using the ambient temperature load.
For the absolute calibration, Orion-KL was observed everyday during the observations, and we adopted main beam temperatures $T_{\rm MB}$ of 75 -- 83\,K from comparison with the data of the KOSMA telescope \citep{1998A&A...335.1049S} and the 60-cm Survey telescope \citep{2007PASJ...59.1005N}.
The pointing accuracy was $\sim 30\arcsec$ as confirmed by observations of Jupiter and N159 (East) ($\alpha_{2000} = 05^{\rm h} 39^{\rm m} 36\fs 08$, $\delta_{2000} = -69\degr 45\arcmin 31\farcs 94$) everyday.
The observations were performed for $\sim$0.7 square degrees with the on-the-fly (OTF) mapping mode.
The data output was at every 30{\arcsec} spacing grid and convolved with a Gaussian function of 45{\arcsec} to the effective angular resolution of 100{\arcsec} (3.6\,pc at a distance of 7.5\,kpc).
The r.m.s.\ noise level after the absolute calibration per channel was $\sim$0.2\,K.

The $^{13}$CO($J$=2--1) (220.399\,GHz) observations were performed for 8 points with the position switching mode in the period from 2009 August to October.
The pointing accuracy $\sim$30{\arcsec} was confirmed by observing Jupiter.
The absolute intensity was scaled by adopting $T_\mathrm{MB}$=17\,K for Ori-KL by comparing with the 60\,cm Survey telescope \citep{2007PASJ...59.1005N}.
The r.m.s.\ noise level was $\sim$0.09\,K.

\subsection{$^{12}$CO($J$=1--0) and $^{13}$CO($J$=1--0) observations with Mopra Telescope}
Observations in the $^{12}$CO($J$=1--0) (115.271\,GHz) and $^{13}$CO($J$=1--0) (110.201\,GHz) lines were carried out by using the Mopra 22\,m telescope of Australia Telescope National Facility (ATNF) from 2008 July to August.
The beam size (HPBW) was 33{\arcsec} at 115\,GHz corresponding to 1.2\,pc at a distance of 7.5\,kpc.
We used the 3\,mm monolithic microwave integrated circuit (MMIC) system whose typical system temperature was 600 -- 1000\,K at the zenith including the atmosphere.
The spectrometer allowed simultaneous observations of $^{12}$CO($J$=1--0) and $^{13}$CO($J$=1--0) by the Mopra Spectrometer (MOPS) zoom band mode, whose frequency resolution was 34\,kHz corresponding to 0.088\,km\,s$^{-1}$ at 115\,GHz.
The spectroscopic data were calibrated to $T_{\rm A}^{\ast}$ by using the ambient temperature load.
$T_{\rm A}^{\ast}$ was corrected to the extended beam temperatures $T_{\rm XB}$ by using extended beam efficiencies $\eta_{\rm XB}$ of 0.60 and 0.52 for $^{12}$CO($J$=1--0) and $^{13}$CO($J$=1--0), respectively, derived by the present observations.
Orion-KL ($\alpha_{2000} = 05^{\rm h} 35^{\rm m} 14\fs 5$, $\delta_{2000} = -5\degr 22\arcmin 29\farcs 6$) was observed everyday as the calibration source, where the extended beam temperatures were 105.5\,K and 18.6\,K in $^{12}$CO($J$=1--0) and $^{13}$CO($J$=1--0), respectively \citep{2005PASA...22...62L}.
The pointing accuracy was achieved to be better than 10{\arcsec} by observing the SiO maser R Car ($\alpha_{2000} = 09^{\rm h} 32^{\rm m} 14\fs 48$, $\delta_{2000} = -62\degr 47\arcmin 19\farcs 7$) every two hours.
The observations were performed for $\sim$0.3 square degrees with the OTF mapping.
The data output was at every 15{\arcsec} spacing grid and convolved with a Gaussian function of 33{\arcsec} to the effective angular resolution of 47{\arcsec} (1.7\,pc at a distance of 7.5\,kpc).
The r.m.s.\ noise level after the absolute calibration per channel were $\sim$0.7\,K and $\sim$0.3\,K in $^{12}$CO($J$=1--0) and $^{13}$CO($J$=1--0), respectively.
The C$^{18}$O($J$=1--0) transition was observed simultaneously but was not detected over the detection limit of 0.3\,K r.m.s.\ noise level.


\section{RESULTS}\label{sec:results}

\subsection{Distribution of the Molecular Gas}

\subsubsection{Global Distribution}\label{subsubsec:global_distribution}
\begin{figure*}
\includegraphics[height=14cm]{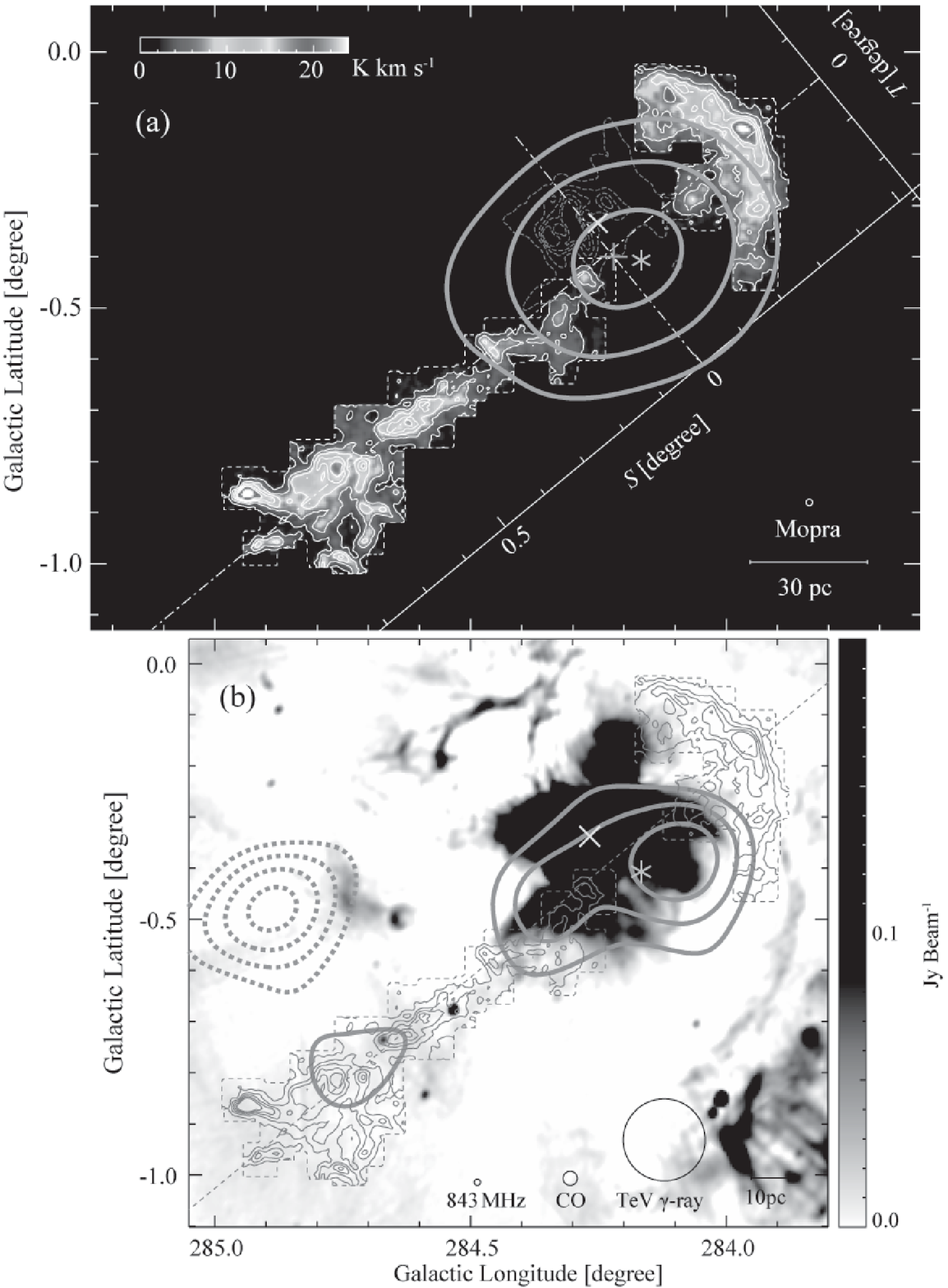}
\caption{
(a) The color image and white contours show the integrated intensity distributions of $^{12}$CO($J$=1--0) in a velocity range of 24 -- 32\,km\,s$^{-1}$.
The white contours are drawn every 3.3\,K\,km\,s$^{-1}$ ($=4\sigma$) up to 5 levels.
The orange contours show significance levels of the HESS J1023$-$575 at a energy range of 0.7 -- 2.5\,TeV started from $7\sigma$ in step of $3\sigma$.
The orange plus is the gravity center of the TeV $\gamma$-ray emission.
The magenta contours show the radio continuum emission at 843\,MHz drawn every 0.5\,Jy\,beam$^{-1}$ from 0.2\,Jy\,beam$^{-1}$.
$S$ axis is taken along the {\it jet} cloud and $T$ axis is taken vertical to $S$ axis (see Section \ref{subsubsec:global_distribution}).
(b) The greyscale image shows the radio continuum emission at 843\,MHz.
The magenta contours show the $^{12}$CO($J$=1--0) (same as Figure \ref{fig1}a).
The orange contours show the TeV $\gamma$-ray smoothed excess image above 2.5\,TeV drawn every 5 excess arcmin$^{-2}$ from 20 excess arcmin$^{-2}$.
The yellow x-marks and green asterisks in the both panels indicate the positions of Wd 2 and PSR J1023$-$5746, respectively.
\label{fig1}}
\end{figure*}

Figure \ref{fig1} shows the distribution of the {\it jet} and {\it arc} clouds toward Wd 2, where the $^{12}$CO($J$=1--0) intensity is integrated in a velocity range from 24\,km\,s$^{-1}$ to 32\,km\,s$^{-1}$.
HESS J1023$-$575 is located between the two clouds.
Paper I suggests that the {\it jet} and {\it arc} clouds are associated with HESS J1023$-5745$.
The Fermi pulsar at 1.8--2.4\,kpc (section \ref{sec:introduction}) is unlikely associated with the {\it jet} and {\it arc} clouds located at 7.5\,kpc (see section \ref{subsec:distance}), and we do not consider that the pulsar is associated with the {\it jet} and {\it arc} clouds.
There is another possible high energy source toward the {\it jet}, the radio continuum point source at ($l$, $b$)=(284{\fdg}55, $-$0{\fdg}70) (Figure \ref{fig1}b), whereas there is no additional support for its association with the {\it jet} cloud (see section \ref{subsec:starformation}).
Therefore we shall hereafter adopt the assumption that the {\it jet} and {\it arc} clouds are associated with the HESS J1023$-$575.
A straight line was determined by an intensity-weighted least-squares fitting to the {\it jet} and {\it arc} clouds and the following relation is obtained;
\begin{equation}
b\,\mbox{(degree)}=(-0.82 \pm 0.01)l\,\mbox{(degree)}+(234.0 \pm 3.3)\,\mbox{(degree)}.
\end{equation}
This line shows a good positional agreement with the geometric center of the HESS J1023$-$575 within 0\fdg 05.
We adopt hereafter the coordinate system defined by the line; $S$ axis is taken along the {\it jet} cloud and $T$ axis is taken vertical to $S$ axis with the origin at $(l,b)=(284\fdg 23, -0\fdg 39)$ close to HESS J1023$-$575.

\begin{figure*}
\includegraphics[height=17cm]{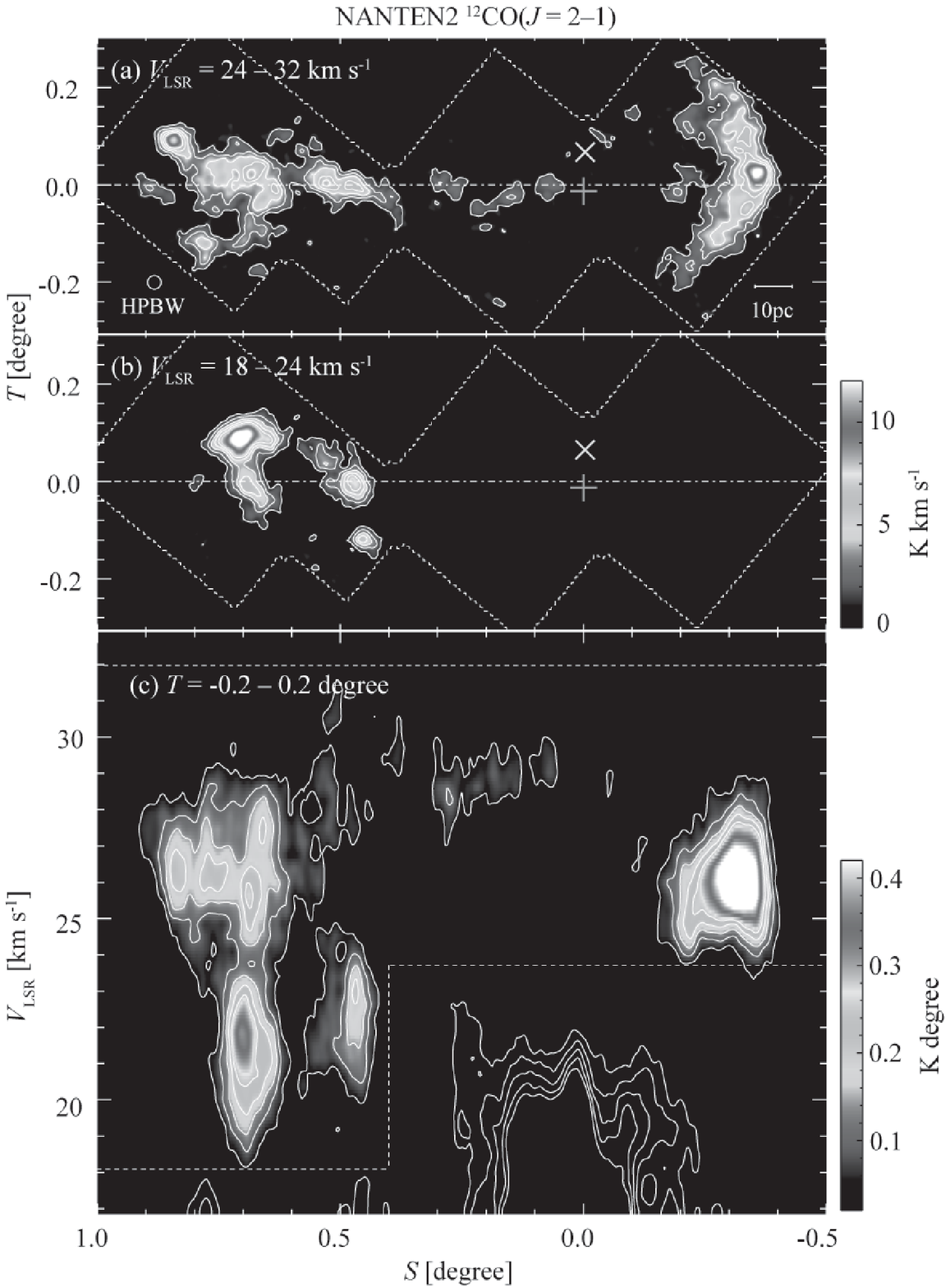}
\caption{
(a) Integrated intensity map of $^{12}$CO($J$=2--1).
The integrated velocity range is 24 -- 32\,km\,s$^{-1}$.
The contours are drawn up to 5 levels every 1.5\,K\,km\,s$^{-1}$ (= $4\sigma$).
The observed region is outlined by dashed lines.
(b) Same as (a) but the integrated range is 18 -- 24\,km\,s$^{-1}$ and the contours are every 1.2\,K\,km\,s$^{-1}$.
(c) Position-velocity diagram integrated along the $T$ axis from $-0\fdg 2$ to $0\fdg2$.
The contours are drawn every 0.059\,K\,degree (= $4\sigma$) up to 5 levels.
The components drawn only with contours correspond to the GMC at 16\,km\,s$^{-1}$ (outside of the dashed-line bouding box).
The pluses and x-marks show the gravity center of HESS J1023$-$575 and center position of Wd 2, respectively.
\label{stmap_12co21}}
\end{figure*}

\begin{figure*}
\includegraphics[height=18cm]{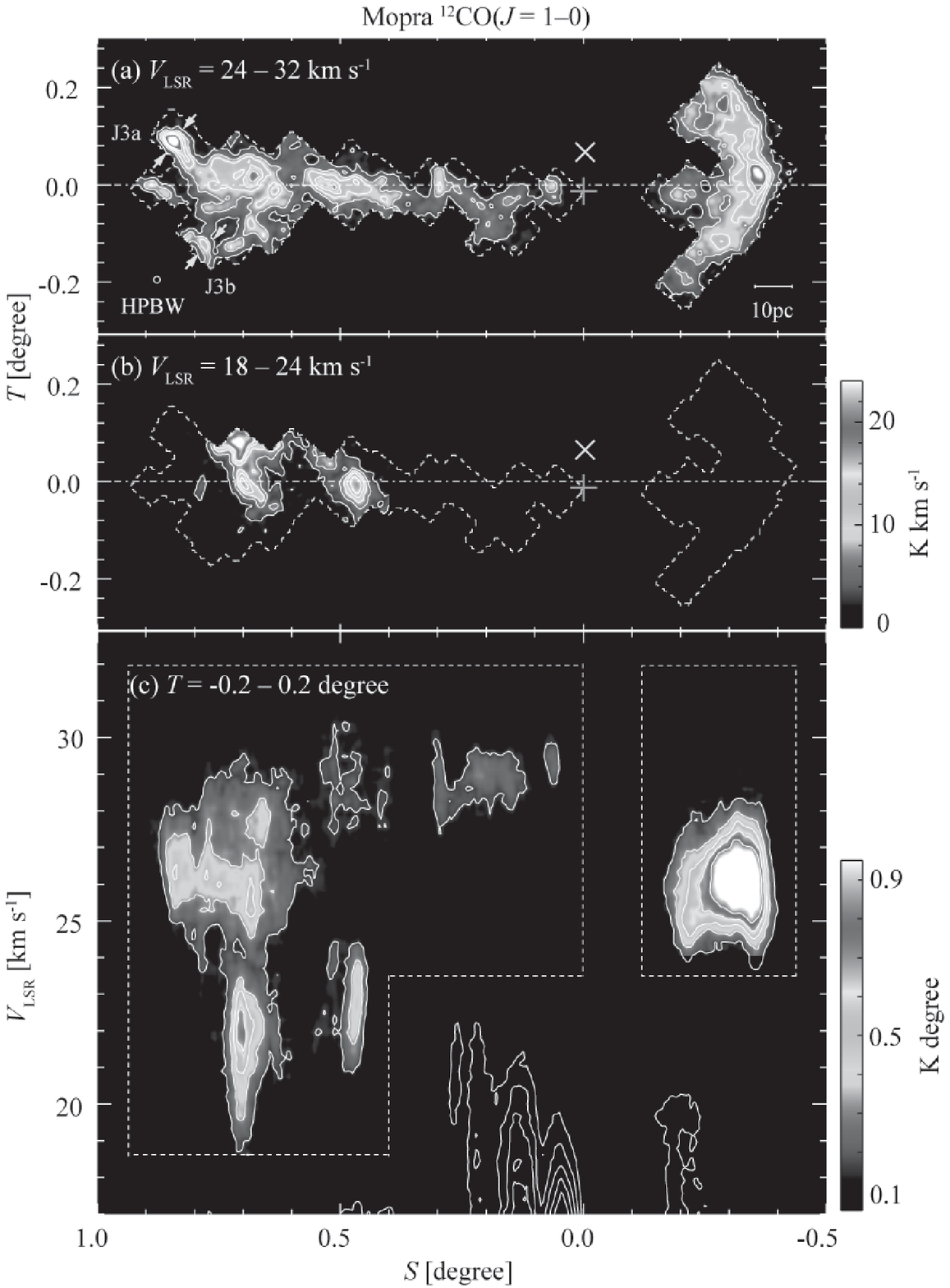}
\caption{
(a) Integrated intensity map of $^{12}$CO($J$=1--0).
The integrated velocity range is 24 -- 32\,km\,s$^{-1}$.
The contours are drawn up to 5 levels every 3.3\,K\,km\,s$^{-1}$ (= $4\sigma$).
The observed region is outlined by dashed lines.
(b) Same as (a) but the integrated range is 18 -- 24\,km\,s$^{-1}$ and the contours are every 2.9\,K\,km\,s$^{-1}$.
(c) Position-velocity diagram integrated along the $T$ axis from $-$0\fdg 2 to 0\fdg 2.
The contours are drawn every 0.16\,K\,degree (= 4$\sigma$) up to 5 levels.
The components drawn only with contours correspond to the GMC at 16\,km\,s$^{-1}$ (outside of the dashed-line bouding box).
The symbols are the same as in Figure \ref{stmap_12co21}.
\label{stmap_12co10}}
\end{figure*}

\begin{figure*}
\includegraphics[height=17cm]{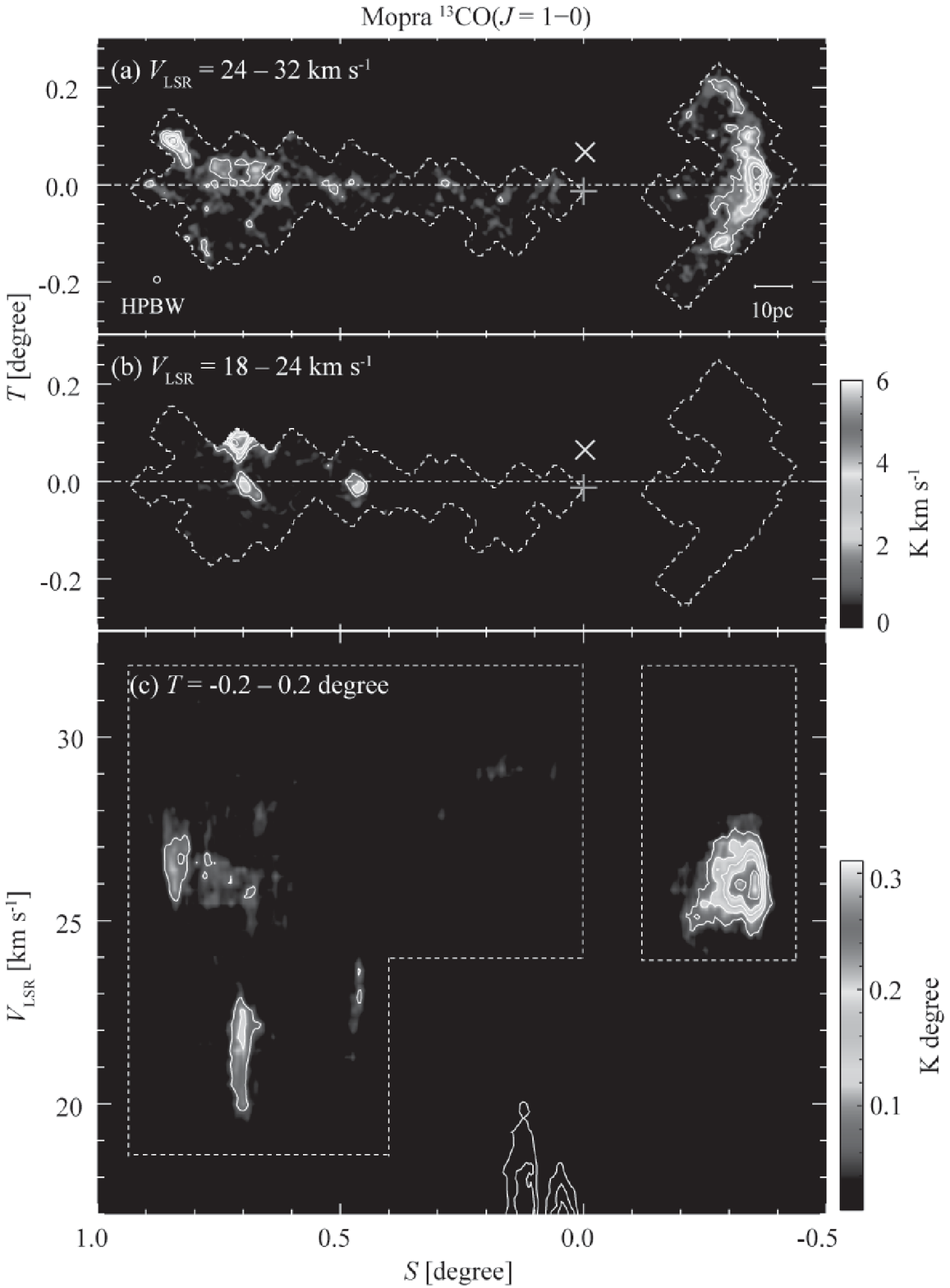}
\caption{
(a) Integrated intensity map of $^{13}$CO($J$=1--0).
The integrated velocity range is 24 -- 32\,km\,s$^{-1}$.
The contours are drawn up to 5 levels every 1.0\,K\,km\,s$^{-1}$ (= 3$\sigma$) from 1.4\,K\,km\,s$^{-1}$ (= 4$\sigma$).
The observed region is outlined by dashed lines.
(b) Same as (a) but he integrated range is 18 -- 24\,km\,s$^{-1}$ and the contours are every 0.9\,K\,km\,s$^{-1}$ from 1.2\,K\,km\,s$^{-1}$.
(c) Position-velocity diagram integrated along the $T$ axis from $-$0\fdg 2 to 0\fdg 2.
The contours are drawn  every 0.20\,K\,km\,s$^{-1}$ (= 3$\sigma$) from 0.26\,K\,km\,s$^{-1}$ (= 4$\sigma$).
The components drawn only with contours correspond to the GMC at 16\,km\,s$^{-1}$ (outside of the dashed-line bouding box).
The symbols are the same as in Figure \ref{stmap_12co21}
\label{stmap_13co10}}
\end{figure*}

Figures \ref{stmap_12co21}, \ref{stmap_12co10} and \ref{stmap_13co10} show detailed distributions of the {\it jet} and {\it arc} clouds in the $^{12}$CO ($J$=2--1, 1--0) and $^{13}$CO ($J$=1--0), respectively.
In each figure, panels (a), (b), and (c) show spatial distribution integrated in the two velocity ranges, 24 -- 32\,km\,s$^{-1}$ and 18 -- 24\,km\,s$^{-1}$, and position($S$)-velocity distribution, respectively.
For the range 18 -- 24\,km\,s$^{-1}$ (each panel b) the $^{12}$ CO ($J$=2--1) data cover the whole distributions while the $J$=1--0 data do not cover them.
In each panel (a) of Figures \ref{stmap_12co21}--\ref{stmap_13co10}, we see winding structures toward $S=$0\fdg 1 -- 0\fdg 3 and $S=$0\fdg 6 -- 0\fdg 9, whose kinematical details are shown in the next section.
The {\it arc} cloud shows a nearly symmetric distribution with respect to the {\it jet} cloud axis.

\begin{figure}
\includegraphics{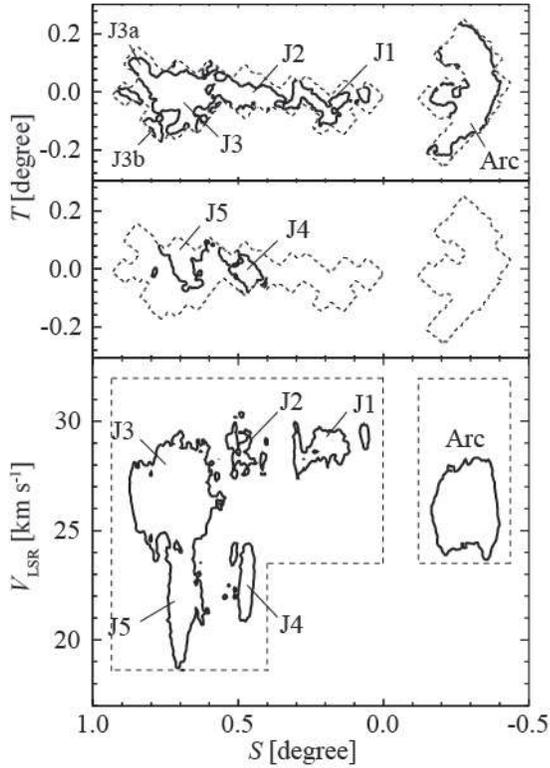}
\caption{
Finding chart for {\it jet} and {\it arc} clouds.
The contours show the  $^{12}$CO($J$=1--0) intensities integrated in the same ranges as Figure \ref{stmap_12co21}.
The contours are drawn at 5$\sigma$ (4.2\,K\,km\,s$^{-1}$ for the top panel and 3.6\,K\,km\,s$^{-1}$ for the middle panel), and at 4$\sigma$ (=0.16\,K\,degree) for the bottom panel. 
\label{finding_chart}}
\end{figure}

Figure \ref{stmap_12co10}, the $^{12}$CO($J$=1--0) data at a higher spatial resolution, was used to define five features J1 -- J5 for the {\it jet} cloud at the lowest contours in Figure \ref{stmap_12co10} (c) and they are depicted in Figure \ref{finding_chart}.
The {\it jet} cloud is divided into the two velocity ranges; the red-shifted features, J1, J2 and J3, and the blue-shifted features, J4 and J5.
As shown in Figures \ref{stmap_12co10} (a) and (b), J2 and J4 show good spatial coincidence, and J3 and J5 as well.
Moreover, J3 and J5 are apparently connected in velocity in Figure \ref{stmap_12co10} (c).
Then, we shall consider two possibilities on the interpretation of J4 and J5.
One is that the blue-shifted and red-shifted features are physically linked.
If this is the case, the velocity span of the {\it jet} cloud becomes as large as $\sim$10\,km\,s$^{-1}$ from the middle to the termination of the {\it jet} cloud. 
Another is that the blue-sifted and red-shifted features are located separately.
If so, velocity widths of the individual features are $\sim$2 -- 5\,km\,s$^{-1}$ (Column 10 in Table \ref{obs_para12co10}).
J3 seems to consist of two branches in the upper and lower parts in $T$.
These two branches seem to bifurcate at $S \sim0\fdg 65$ (Figures \ref{stmap_12co10} a and \ref{stmap_13co10} a) and have peaks at each eastern terminal positions named J3a and J3b (Figure \ref{finding_chart}).
The {\it arc} cloud is in a velocity range from 24\,km\,s$^{-1}$ to 29\,km\,s$^{-1}$.

\subsubsection{Details of the {\it Jet} Cloud}\label{subsubsec:details_jet}
\begin{figure*}
\includegraphics{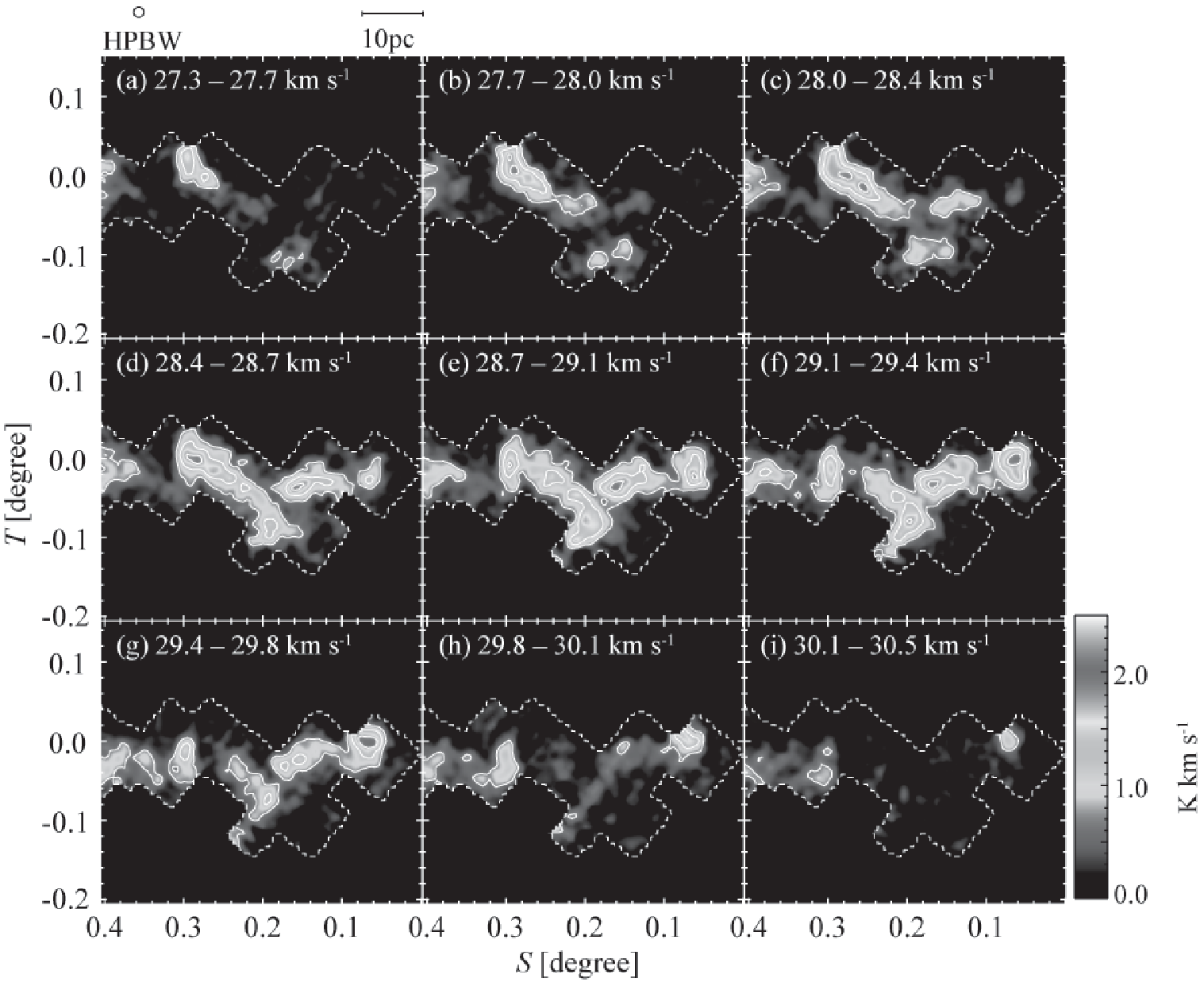}
\caption{
Velocity channel maps of J1 in $^{12}$CO($J$=1--0).
The integrated velocity channel per map is 4 corresponding to 0.35\,km\,s$^{-1}$.
The contours are drawn every 0.70\,K\,km\,s$^{-1}$ (=4$\sigma$).
The observed region is outlined by dashed lines.
\label{chmap_j1}}
\end{figure*}

\begin{figure*}
\includegraphics{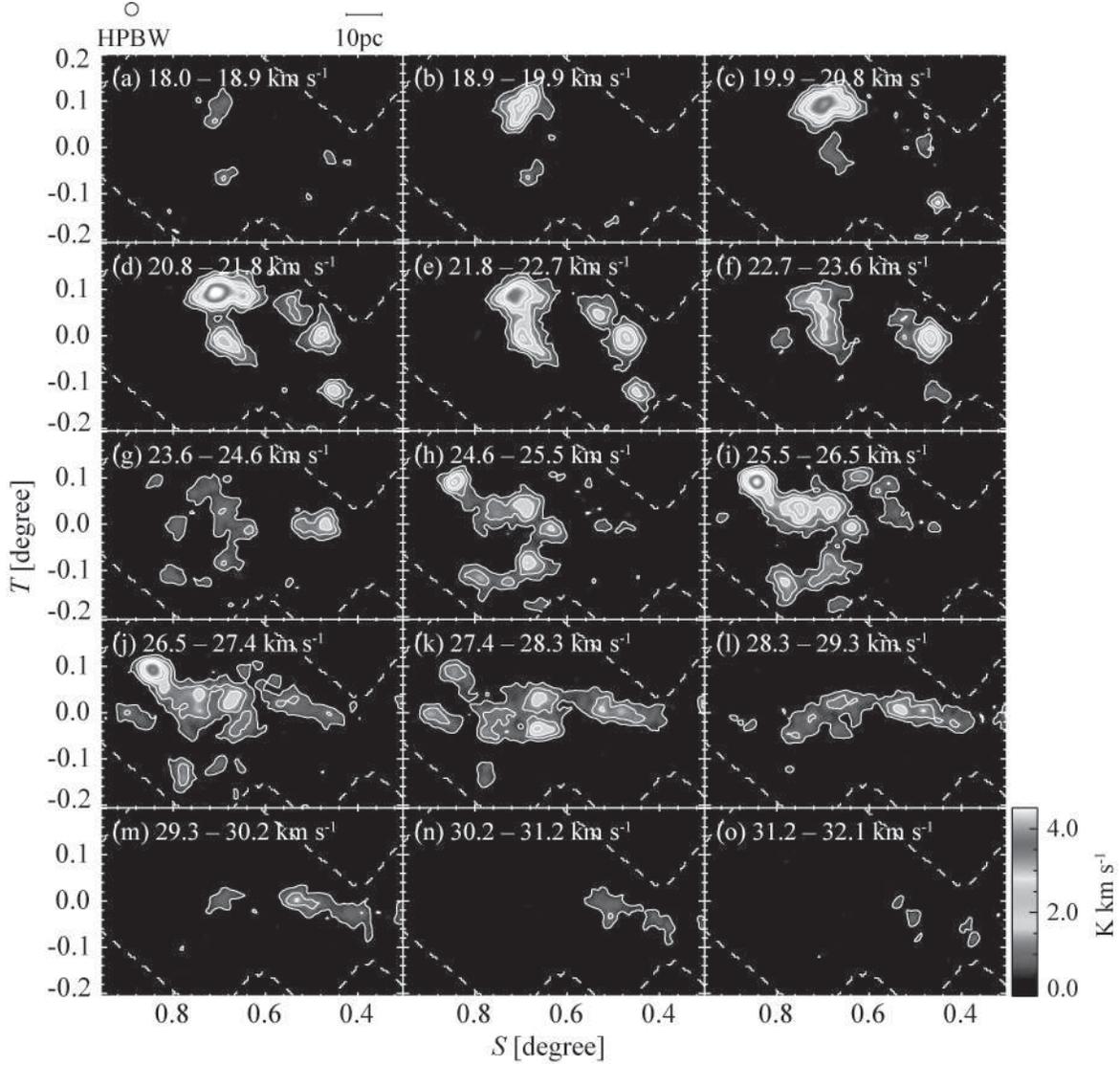}
\caption{
Velocity channel maps of J2 -- J5 in $^{12}$CO($J$=2--1).
The integrated velocity channel per map is 5 corresponding to 0.94\,km\,s$^{-1}$.
The contours are drawn every 0.50\,K\,km\,s$^{-1}$ (=4$\sigma$)  up to 5 levels.
The observed region is outlined by dashed lines.
\label{chmap_j2j3}}
\end{figure*}

Figure \ref{chmap_j1} shows the velocity channel distributions of J1 in $^{12}$CO($J$=1--0).
The component in the middle of J1 in $S=$ 0{\fdg}1 -- 0{\fdg}30 is winding and shows a velocity gradient from $S=$ 0{\fdg}3 to $S=$ 0{\fdg}1.
J1 appears to show a perpendicular component at $S=$0{\fdg}3 in Figures \ref{chmap_j1} (e) -- (h).
Figure \ref{chmap_j2j3} shows the velocity channel distributions of J2 -- J5 in $^{12}$CO($J$=2--1).
The bifurcation to the upper and lower features in J3 are seen in panels (h) and (i).
The upper branch shows bending at $(S, T)=(0{\fdg}8, 0{\fdg}05)$ and the lower branch also shows bending at ($S$, $T$)=(0{\fdg}7, $-$0{\fdg}1).

\subsubsection{Details of the {\it Arc} Cloud and Evidence for the Counter {\it Jet} Cloud}\label{subsubsec:details_arc}
\begin{figure*}
\includegraphics{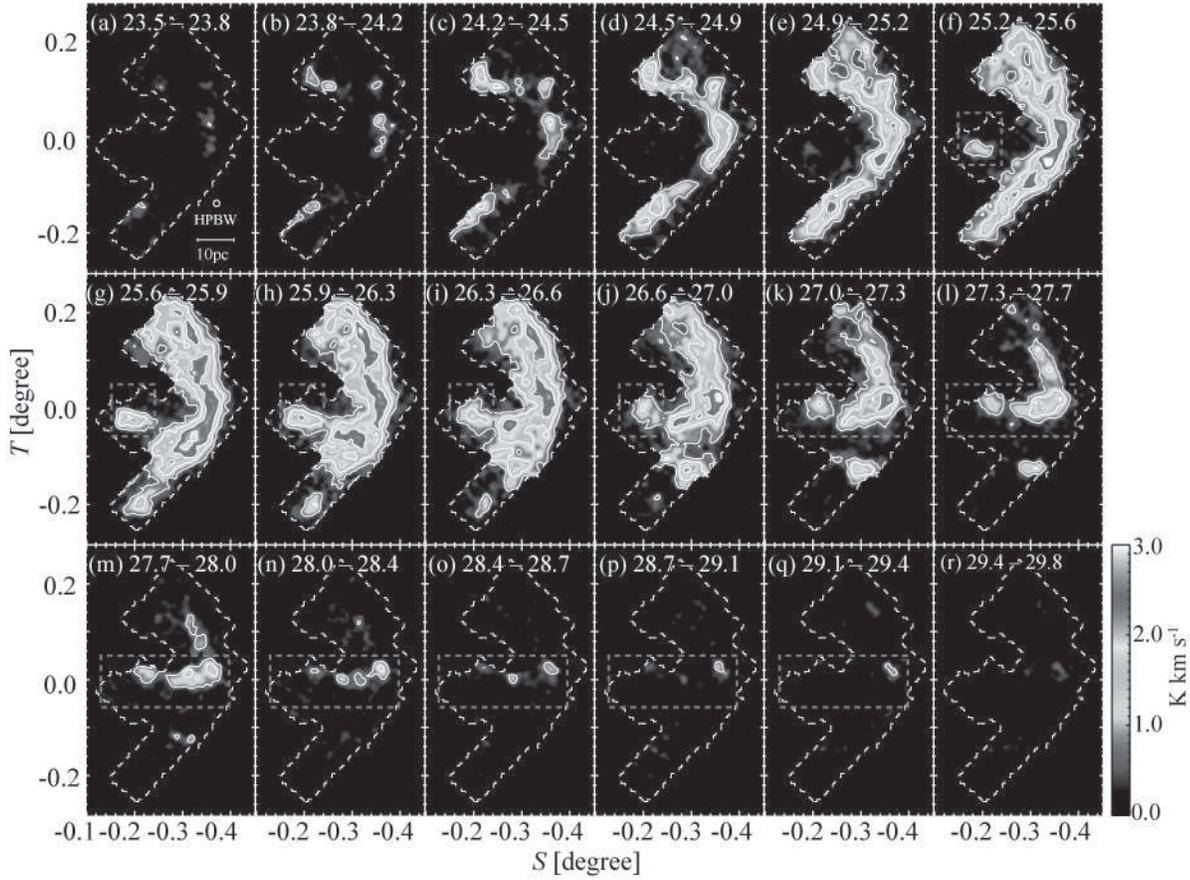}
\caption{
Velocity channel maps of arc in $^{12}$CO($J$=1--0).
The integrated velocity channel per map is 4 corresponding to 0.35\,km\,s$^{-1}$.
The contours are drawn every 0.70\,K\,km\,s$^{-1}$ (=4$\sigma$).
The thick line boxes (magenta in a color version) are areas where the components of the western {\it jet} cloud are extracted as mentioned in Section \ref{subsubsec:details_arc}.
The observed region is outlined by dashed lines.
\label{chmap_arc}}
\end{figure*}

Figure \ref{chmap_arc} shows velocity channel distributions of the {\it arc} cloud in $^{12}$CO($J$=1--0).
The integrated intensity in Figure 1 shows a clear crescent shape but the shape varies significantly in the individual velocity channels; e.g., the distribution in Figure \ref{chmap_arc} (d) is bent sharply.
Another new aspect is that a feature elongated along the {\it jet} cloud axis is clearly seen in the velocity range from 25.2\,km\,s$^{-1}$ to 29.4\,km\,s$^{-1}$.
This feature is seen in $S=-$0{\fdg}4 -- $-$0{\fdg}2, showing a velocity gradient from 29.4\,km\,s$^{-1}$ to 25.2\,km\,s$^{-1}$ with $S$.
We tentatively interpret this feature as a counterpart to the previously known {\it jet} cloud and shall call ``western {\it jet} cloud'' for convenience.
In a similar manner, we call the {\it jet} cloud in the east of Wd 2 ``eastern {\it jet} cloud'' hereafter.

\begin{figure*}
\includegraphics{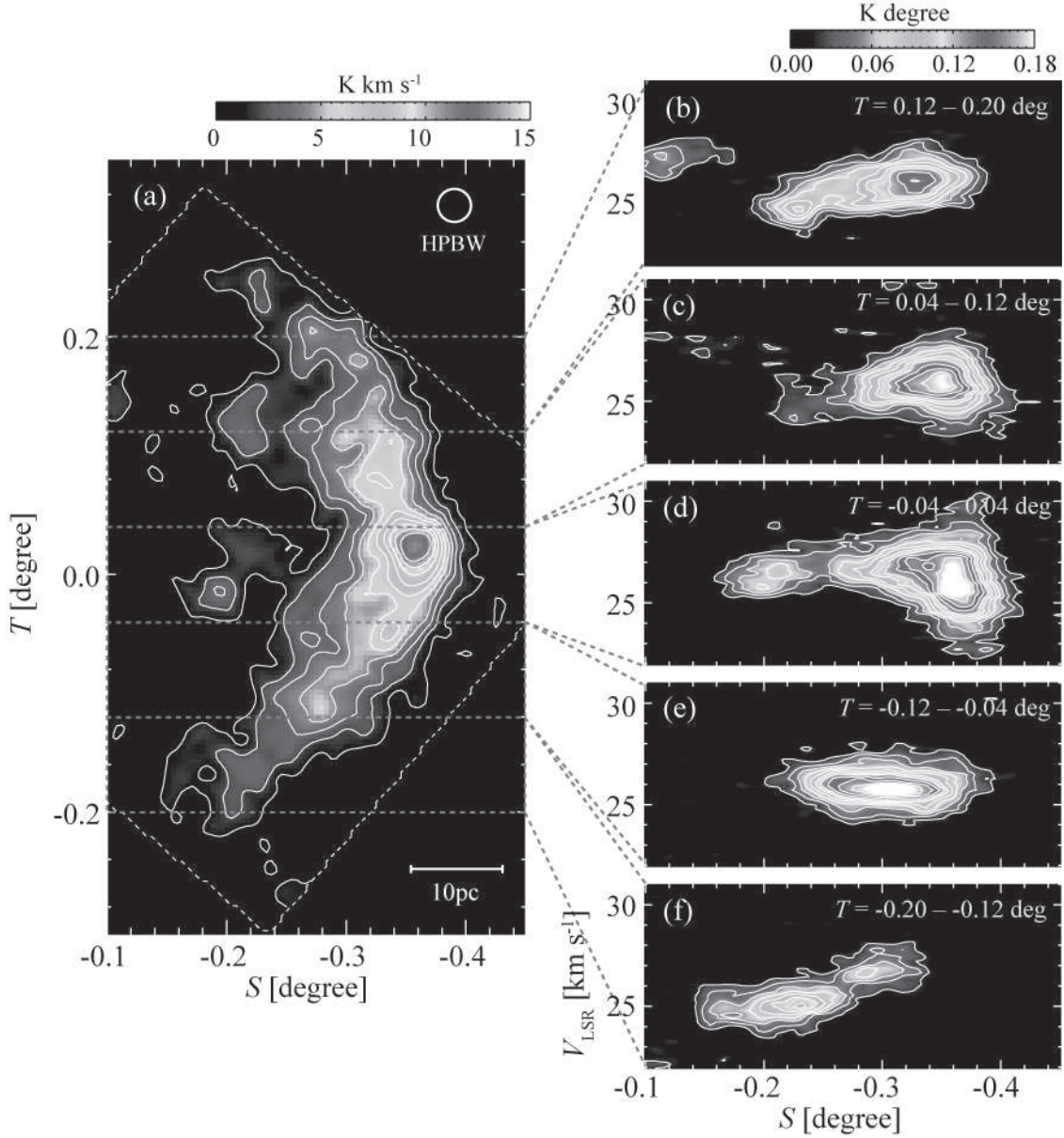}
\caption{
Position-velocity diagrams of {\it arc} integrated in various ranges of $T$.
(a) Spatial intensity map of the {\it arc} cloud in $^{12}$CO($J$=2--1) integrated in 24 -- 32\,km\,s$^{-1}$.
The contours are drawn every 1.5\,K\,km\,s$^{-1}$ (=4$\sigma$).
The thick line boxes (magenta in a color version) indicate the integrated regions of $T$ in the panels (b) -- (f).
(b) -- (f) Position-velocity diagrams integrated along the direction of $T$.
The contours are drawn every 0.016\,K\,degree (=3$\sigma$) from 0.022\,K\,degree (=4$\sigma$).
\label{arc_pvmap}}
\end{figure*}

The western edge of the western {\it jet} cloud coincides with the peak of the {\it arc} cloud.
Figure \ref{arc_pvmap} shows more details of the {\it arc} cloud in the position-velocity diagrams.
All the diagrams show small linewidth of less than 3\,km\,s$^{-1}$ and show no systematic velocity shift in position.
If the {\it arc} cloud is part of an expanding shell as suggested by its crescent shape, we expect a systematic change of velocity with the projected radius from the center.
We suggest that the {\it arc} cloud may not be a shell but is a thin filamentary feature.
In the appendix \ref{sec:expandingshell}, a schematic view of a typical position-velocity diagram for an expanding ellipsoidal shell is shown for comparison.

\begin{figure*}
\includegraphics{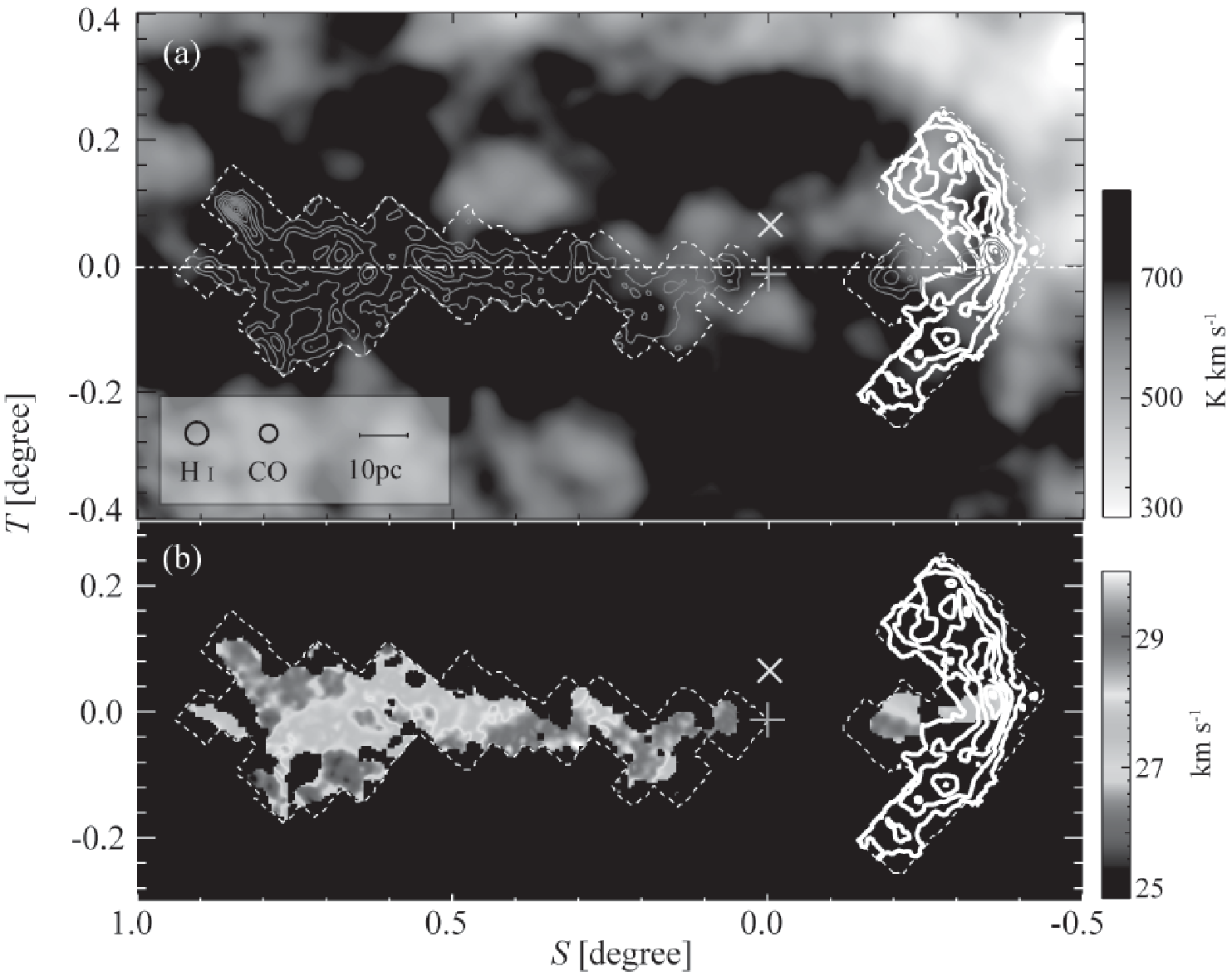}
\caption{
(a) Integrated intensities of eastern and western {\it jet} clouds in $^{12}$CO($J$=1--0) ({\it magenta contours}).
The gray scale shows the integrated intensities of \ion{H}{1} 21\,cm line \citep{2005ApJS..158..178M}.
The white contours show the integrated intensity of the {\it arc} cloud in $^{12}$CO($J$=1--0).
All the data are integrated in a range of 24 -- 32\,km\,s$^{-1}$.
The western {\it jet} cloud is extracted from the magenta boxes in Figure \ref{chmap_arc}.
The all contours are drawn every 3.3\,K\,km\,s$^{-1}$ (=4$\sigma$).
(b) Intensity-weighted first moment map of the eastern and western {\it jet} clouds.
The contours are the same as in the panel (a).
The symbols are the same as in Figure \ref{stmap_12co21}.
\label{jet}}
\end{figure*}

Figure \ref{jet} (a) shows the distribution of the eastern and western {\it jet} clouds and the {\it arc} cloud superposed on the \ion{H}{1} distribution in the same velocity range from the Southern Galactic Plane Survey \citep{2005ApJS..158..178M}.
Figure \ref{jet} (b) shows the intensity-weighted first moment map which indicates the barycentric velocity distribution in the eastern and western {\it jet} clouds and we see that the barycentric velocity increases from 26\,km\,s$^{-1}$ to 29\,km\,s$^{-1}$ in the eastern {\it jet} cloud and decreases from 28\,km\,s$^{-1}$ to 26\,km\,s$^{-1}$ in the western {\it jet} cloud from the center position of HESS J1023$-$575.
The alignment of the two {\it jet} clouds seems good while the lengths of them are asymmetric; the eastern {\it jet} cloud has 100\,pc in length and the western {\it jet} cloud has 40\,pc in length at a distance of 7.5\,kpc.
We see a sign of an \ion{H}{1} shell surrounding HESS J1023$-$575 as recognized in Paper I.
The shell is elongated along the {\it jet} cloud axis with a size of 45\,pc $\times$ 32\,pc (Figure \ref{jet} a).
Paper I suggests that the {\it arc} cloud is part of the \ion{H}{1} shell formed under a compression related to HESS J1023$-$575, since HESS J1023$-$575 appears to be located toward the \ion{H}{1} depression in the shell.
We also confirm the elongated \ion{H}{1} emission toward the eastern {\it jet} cloud (Paper I).

\subsection{Distance of the {\it Jet} and {\it Arc} Clouds}\label{subsec:distance}

\begin{figure}
\includegraphics{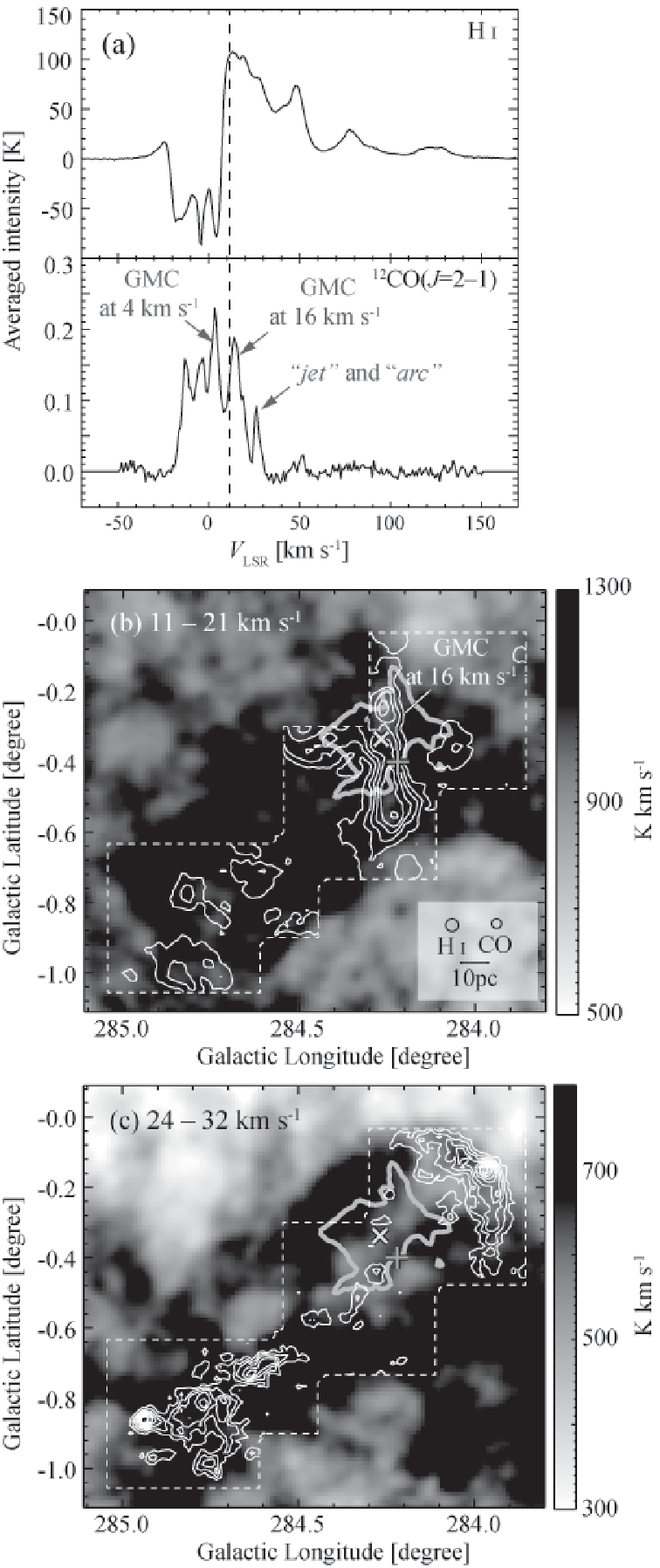}
\caption{
(a) Profiles of \ion{H}{1} 21\,cm ({\it top}) \citep{2005ApJS..158..178M} and $^{12}$CO($J$=2--1) ({\it bottom}) lines toward RCW 49.
The profiles are made by averaging in ($l$, $b$)=(284{\fdg}2 -- 284{\fdg}4, $-$0{\fdg}4 -- $-$0{\fdg}25) for \ion{H}{1} and in ($l$, $b$)=(283{\fdg}8 -- 285{\fdg}1, $-$1{\fdg}1 -- 0{\fdg}0) for CO.
Components of GMCs at the 4\,km\,s$^{-1}$ and 16\,km\,s$^{-1}$, and the {\it jet} and {\it arc} clouds are indicated in the CO profile.
The dashed line shows the most redshifted velocity where the absorption occurs.
(b) Integrated intensity map of \ion{H}{1} 21\,cm line in the velocity range of the GMC at 16\,km\, s$^{-1}$.
The thin contours show the $^{12}$CO($J$=2--1) integrated intensities drawn every 5.8\,K\,km\,s$^{-1}$ (=15$\sigma$) from 3.1\,K\,km\,s$^{-1}$ (=8$\sigma$).
(c) Integrated intensity map of \ion{H}{1} 21\,cm line in the velocity range of the {\it jet} and {\it arc} clouds.
The thin contours are drawn every 1.5\,K\,km\,s$^{-1}$ (=4$\sigma$).
The symbols in panels (b) and (c) are the same as in Figure \ref{stmap_12co21}, and the thick contours (cyan in a color version) outline RCW49.
\label{h1_abs}}
\end{figure}

In order to test the distance of the relevant CO features, we show in Figure \ref{h1_abs} (a) a comparison between the CO and \ion{H}{1} line profiles.
The \ion{H}{1} should absorb the radio continuum of RCW 49 if the gas is in the foreground.
The \ion{H}{1} profile in Figure \ref{h1_abs} (a) shows that only the blue-shifted features show signs of absorption toward the radio source, whereas the more red-shifted \ion{H}{1} features show no sign of absorption (Figures \ref{h1_abs} b and c).
So, the GMC at 4\,km\,s$^{-1}$ is located in front of RCW 49 and the GMC at 16\,km\,s$^{-1}$ and the {\it jet} and {\it arc} clouds are behind RCW 49.
There is no further hint about the physical linkage between the Wd 2-RCW 49-GMCs system and the {\it jet}-{\it arc} cloud system.
Here we tentatively assume that the two systems are not physically associated.
This assumption implies that HESS J1023$-$575 is not physically connected with Wd 2.
We adopt a kinematic distance of the {\it jet} and {\it arc} clouds of $7.5_{-0.5}^{+0.2}$\,kpc from the averaged intensity weighted velocity of 26.5\,km\,s$^{-1}$ by using a flat-rotation model \citep{1993A&A...275...67B}, which is 2.1\,kpc greater than that of the Wd 2-RCW 49-GMCs system.
Here we assume that the galactic distance and rotation velocity at the Sun are 8.5\,kpc and 220\,km\,s$^{-1}$.
The superscript and subscript values mean uncertainty of the distance which are derived by the kinematic distance of the most red-shifted and blue-shifted clouds among the {\it arc} and {\it jet} clouds in Table \ref{obs_para12co10}. 

\subsection{Physical Parameters of the {\it Jet} and {\it Arc} Clouds}

\subsubsection{The Radius, Mass, and Age of the Clouds}\label{subsubsec:radius}
Radii and masses of the {\it arc} cloud and J1 -- J5 at a distance of 7.5\,kpc are summarized in Table \ref{phys_param}.
A radius $r$ of each cloud is derived by
\begin{equation}
r = \sqrt{\frac{A}{\pi} - { \left( \frac{\theta_{\rm HPBW}}{2} \right)}^2} \mbox{,} 
\end{equation}
where $\theta_{\rm HPBW}$ is the beam size in pc.
$A$ is area where $^{12}$CO($J$=1--0) is detected above 4$\sigma$ in the integrated intensity in a range of 24 -- 32\,km\,s$^{-1}$ (J1 -- J3) or $^{12}$CO($J$=2--1) in a range of 18 -- 24\,km\,s$^{-1}$ (J4 and J5; they are not fully covered in $J$=1--0).
Mass is derived by
\begin{equation}
M = \mu m_{\rm H} \sum_i D^2 \Omega N_i(\mathrm{H_2}),
\end{equation}
where $\mu$ is the mean molecular weight per H molecule, $m_{\rm H}$ the proton mass, $D$ the distance, $\Omega$ the solid angle of one pixel, and $N_i(\rm H_2)$ the column density of each pixel.
We use $\mu=2.8$ by adopting the helium abundance of 20\% to the molecular hydrogen.
The column density is estimated by using a conversion factor called as {\it X}-factor $X_{\rm CO}$ from the integrated intensity of $^{12}$CO($J$=1--0), $W(\rm CO)$, to the column density as $N({\rm H_2}) = X_{\rm CO}\cdot W({\rm CO})$.
For J4 and J5 we converted $^{12}$CO($J$=2--1) integrated-intensities into $^{12}$CO($J$=1--0) intensities with $J$=1--0/$J$=2--1 ratios of 0.43 (J4) or 0.47 (J5).
In the present work we adopt $1.6 \times 10^{20}$\,cm$^{-2}$\,(K\,km\,s$^{-1}$)$^{-1}$ as the {\it X}-factor \citep{1997ApJ...481..205H} obtained by {\rm EGRET} observations instead of $2.0 \times 10^{20}$\,cm$^{-2}$\,(K\,km\,s$^{-1}$)$^{-1}$ \citep{1993ApJ...416..587B} used in Paper I, since the former was derived with CO data by taking into account a correction factor of 1.22 in the calibration \citep{1997ApJ...481..205H}.
 We note that values of the {\it X}-factor derived by various works based on $\gamma$-ray observations, infrared observations, and so on \citep[e.g.,][]{1986A&A...154...25B, 1998ApJ...507..507R, 2001ApJ...547..792D} are in a range of $(1\mbox{--}3) \times 10^{20}$\,cm$^{-2}$\,(K\,km\,s$^{-1}$)$^{-1}$.
Therefore, the obtained molecular masses may include $\sim$50 \% uncertainty.

From Table \ref{phys_param} the molecular mass of the eastern {\it jet} cloud (EJ), that of the western {\it jet} cloud (WJ), and that of the {\it arc} cloud are estimated to be $M({\rm EJ}) = 5.6 \times 10^4 M_\sun$, $M({\rm WJ})=2.9 \times 10^3 M_\sun$, and $M({\rm arc}) = 2.5 \times 10^4 M_\sun$, respectively.
The \ion{H}{1} mass for an area $1\fdg 5 \times 0\fdg 8$ shown in Figure \ref{jet} (a) amounts to $1.5 \times 10^5 M_\sun$ and the molecular gas corresponds to $\sim$60\% of the atomic mass.
These values differ from those in Paper I mainly because of the distance revised and the different {\it X}-factor.

Crossing timescale of a molecular cloud obtained by dividing the molecular cloud size by the linewidth is often used as a measure of typical cloud timescale, since this means a time in which the cloud can significantly alter its morphology.
The width of the {\it jet} and {\it arc} clouds is $\sim$5 -- 10\,pc, and their linewidths $\Delta V_\mathrm{comp}$ are 2 -- 5\,km\,s$^{-1}$ (Table \ref{obs_param}).
The crossing timescale of the {\it jet} and {\it arc} clouds is estimated to be in the order of a few Myr.

\subsubsection{The Line Intensity Ratios; the $^{12}$CO ($J$=2--1, 1--0) and $^{13}$CO ($J$=1--0) Transitions}\label{subsubsec:intensity_ratio}
\begin{figure*}
\includegraphics{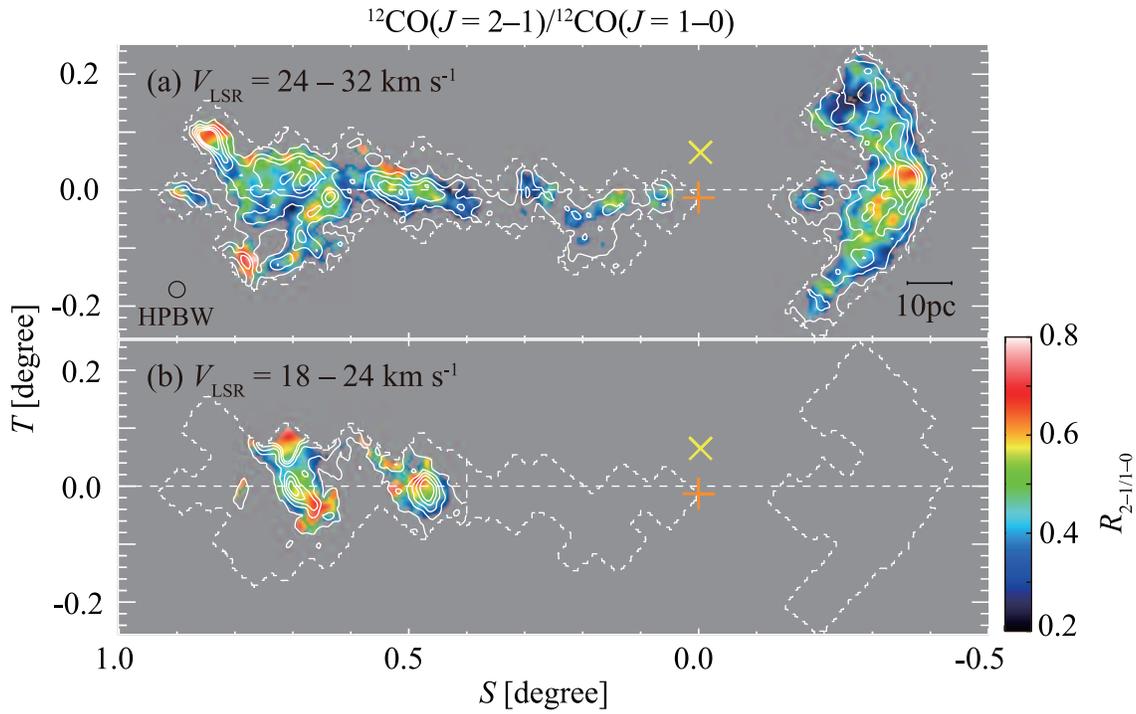}
\caption{
Integrated intensity ratio maps of $^{12}$CO($J$=2--1) to $^{12}$CO($J$=1--0), $R_{\rm 2-1/1-0}$.
The integrated ranges are written in the panels.
Only pixels where the integrated intensities exceeding 4$\sigma$ in both of the lines are shown.
(a) The contours are the $^{12}$CO($J$=1--0) integrated intensities every 3.3\,K\,km\,s$^{-1}$ (=4$\sigma$) up to 4 levels.
(b) The contours are the $^{12}$CO($J$=1--0) integrated intensities every 2.9\,K\,km\,s$^{-1}$ (=4$\sigma$) up to 4 levels.
The symbols are the same as in Figure \ref{stmap_12co21}.
\label{ratio_2-1_1-0}}
\end{figure*}

\begin{figure*}
\includegraphics{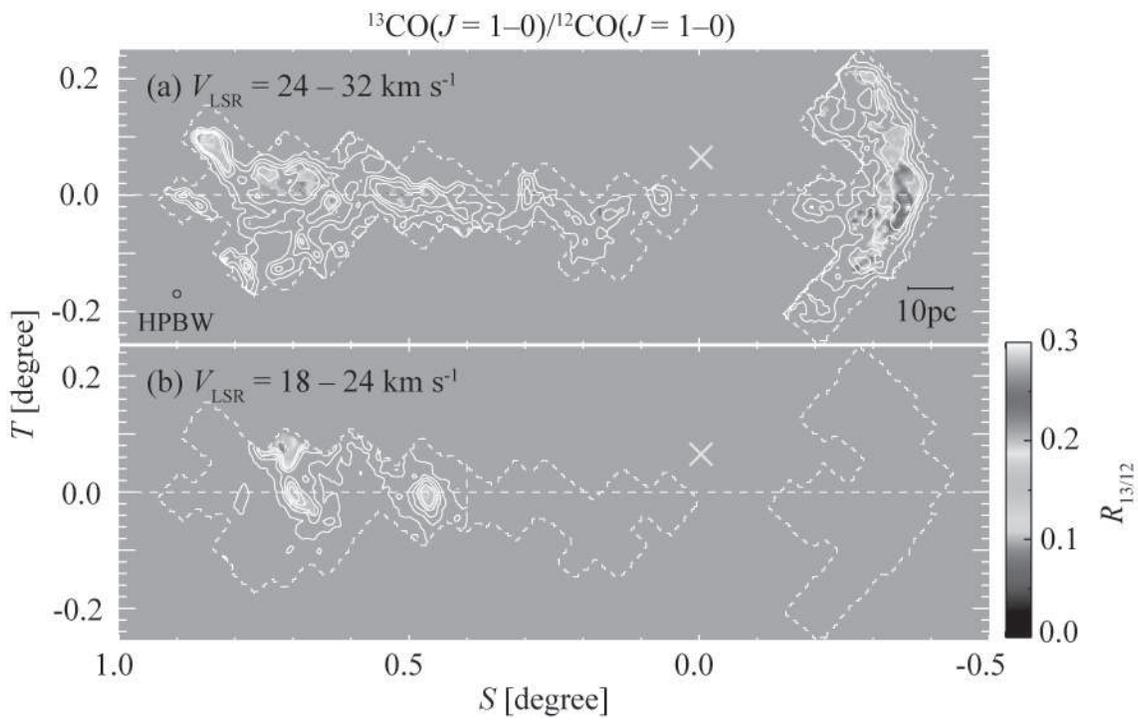}
\caption{
Integrated intensity ratio maps of $^{13}$CO($J$=1--0) to $^{12}$CO($J$=1--0), $R_{\rm 13/12}$.
The integrated ranges are written in the panels.
Only pixels where the integrated intensities exceeding 4$\sigma$ in both of the lines were shown.
The contours and symbols are the same as in Figure \ref{ratio_2-1_1-0}.
\label{ratio_12_13}}
\end{figure*}

Ratios between different $J$ transitions and isotopes reflect excitation conditions and density in the molecular clouds.
The excitation of molecular transitions depends on the temperature and density in the molecular gas.
The line intensity ratio of the three transitions are summarized in Figures \ref{ratio_2-1_1-0} and \ref{ratio_12_13}.
The 1--0 data were Gaussian smoothed to a 90{\arcsec} beam size.
Figure \ref{ratio_2-1_1-0} shows distributions of the integrated intensity ratio of $^{12}$CO($J$=2--1) to $^{12}$CO($J$=1--0), $R_{2-1/1-0}$, where the both lines are detected above $4\sigma$.
Although the average of the ratio is 0.45, the ratio exceeds 0.6 at the peak position of the {\it arc} cloud, at J3a and J3b, and at the peak positions of J4 and J5.
The ratio is high at the edges of the clouds in some places, which could be artifacts due to low S/N ratios.
Figure \ref{ratio_12_13} shows spatial distributions of the integrated intensity ratio of $^{13}$CO($J$=1--0) to $^{12}$CO($J$=1--0), $R_{13/12}$, where both lines are detected above 4$\sigma$.
$R_{13/12}$ reflects molecular column density distributions.
The ratio is enhanced to be around 0.25 in the {\it arc} cloud, J3a and the two peaks of J5.
It implies the column density becomes relatively higher at these positions than the rest.

\subsubsection{The LVG Analysis}\label{subsubsec:lvg}
In order to estimate kinetic temperature and number density of the molecular clouds, we carried out the LVG (large velocity gradient) analysis \citep{1974ApJ...189..441G} for points where the $^{13}$CO($J$=2--1) pointed observations were made.
The employed model assumes a spherically symetric cloud where kinetic temperature $T_{\rm kin}$, number density $n({\rm H}_2)$ and radial velocity gradient $dv/dr$ are taken to be uniform.
We changed $T_{\rm kin}$ and $n({\rm H}_2)$ within $T_{\rm kin} = 6$ -- 500\,K and $n({\rm H}_2) = 10$ -- $10^6$\,cm$^{-3}$, where we fix $X/(dv/dr)$ for a cloud size of a few pc and a linewidth of a few km\,s$^{-1}$.
For an isotope ratio of [$^{12}$C]/[$^{13}$C], we adopt 75 at the Galacto-centric distance of $\sim$9\,kpc \citep{2005ApJ...634.1126M}.

\begin{figure*}
\includegraphics{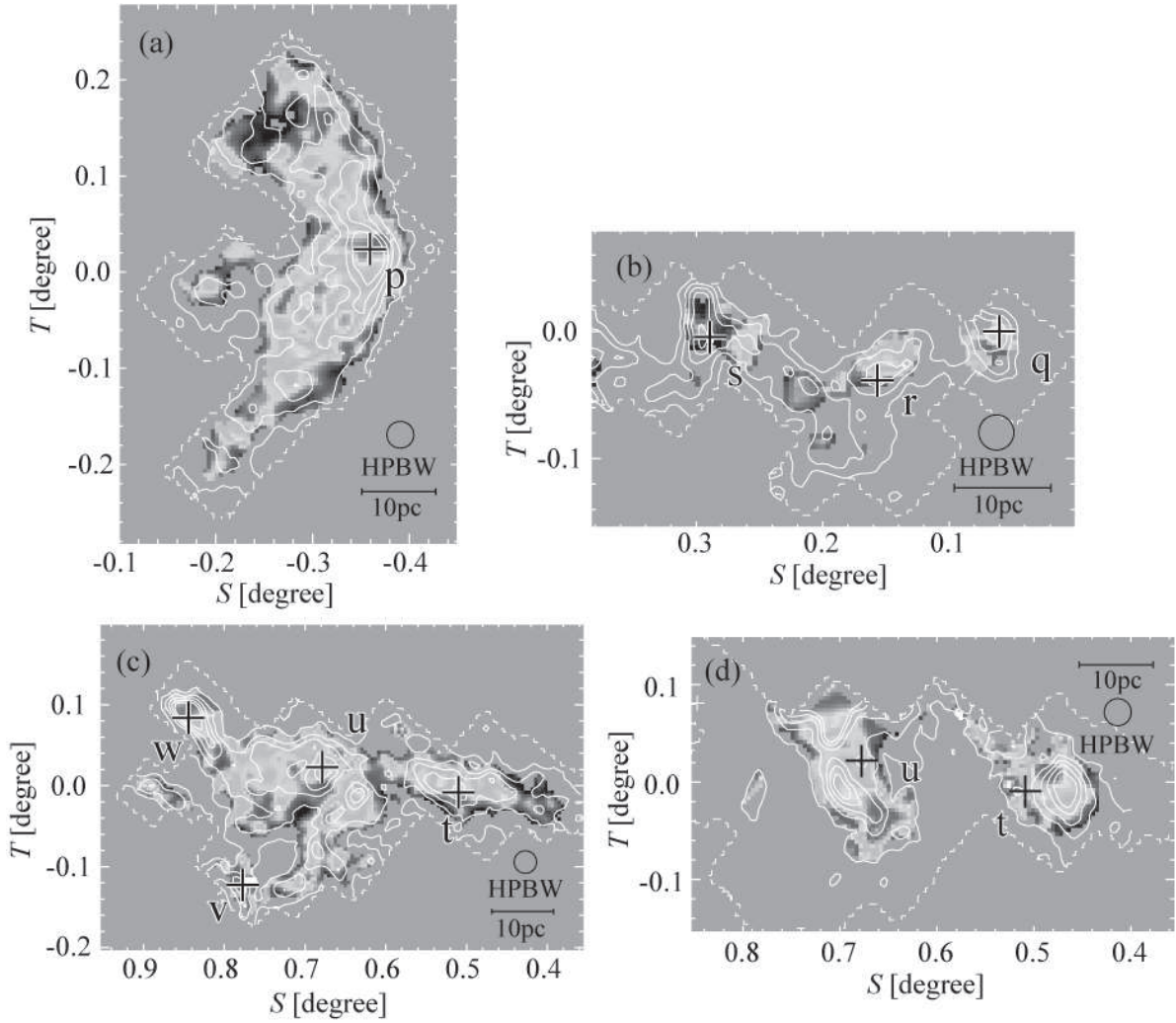}
\caption{
Closeup of Figure \ref{ratio_2-1_1-0}. The pluses point positions where the LVG analyses are performed. 
\label{ratio_2-1_1-0c}}
\end{figure*}

\begin{figure*}
\includegraphics{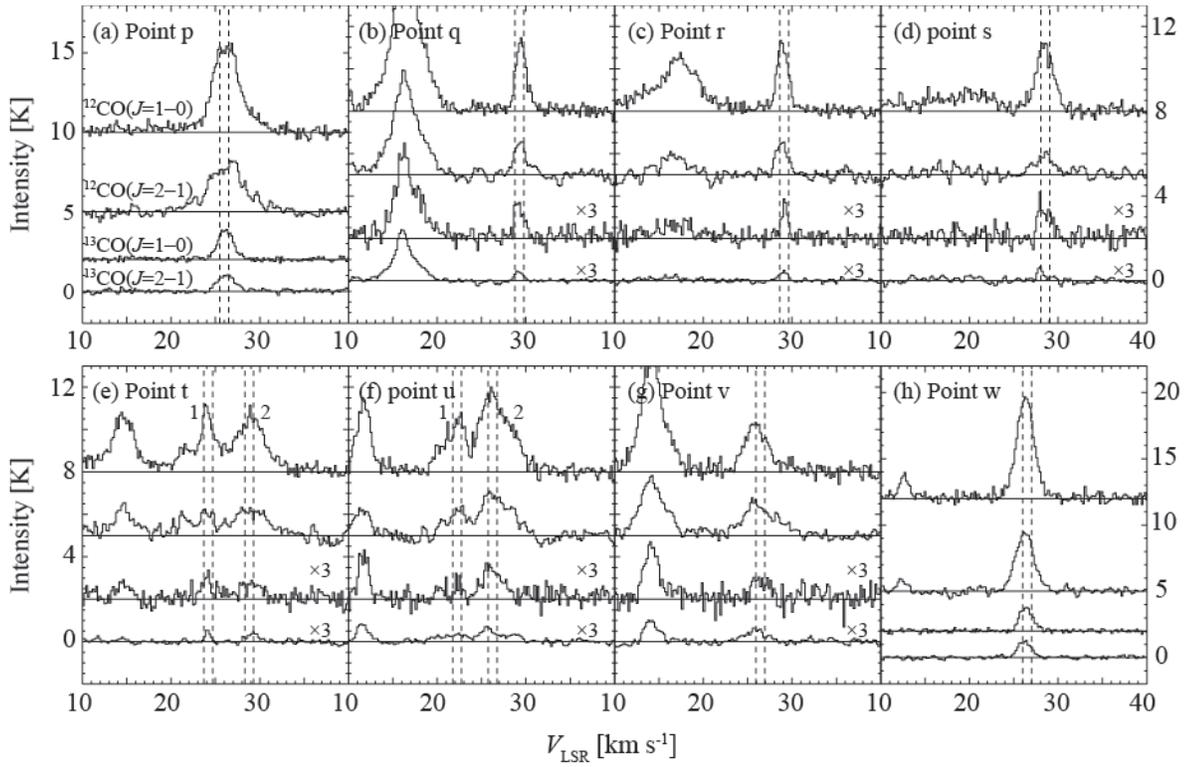}
\caption{
Spectra of $^{12}$CO($J$=1--0), $^{13}$CO($J$=1--0), $^{12}$CO($J$=2--1), and $^{13}$CO($J$=2--1) where LVG analyses are performed.
The $^{13}$CO($J$=1--0) and $^{13}$CO($J$=2--1) spectra in (b) -- (g) are multiplied by a factor of 3.
The vertical broken lines show the integrated range of 1\,km\,s$^{-1}$ around the barycentric velocities used in the LVG analyses.
\label{spectra}}
\end{figure*}

\begin{figure*}
\includegraphics{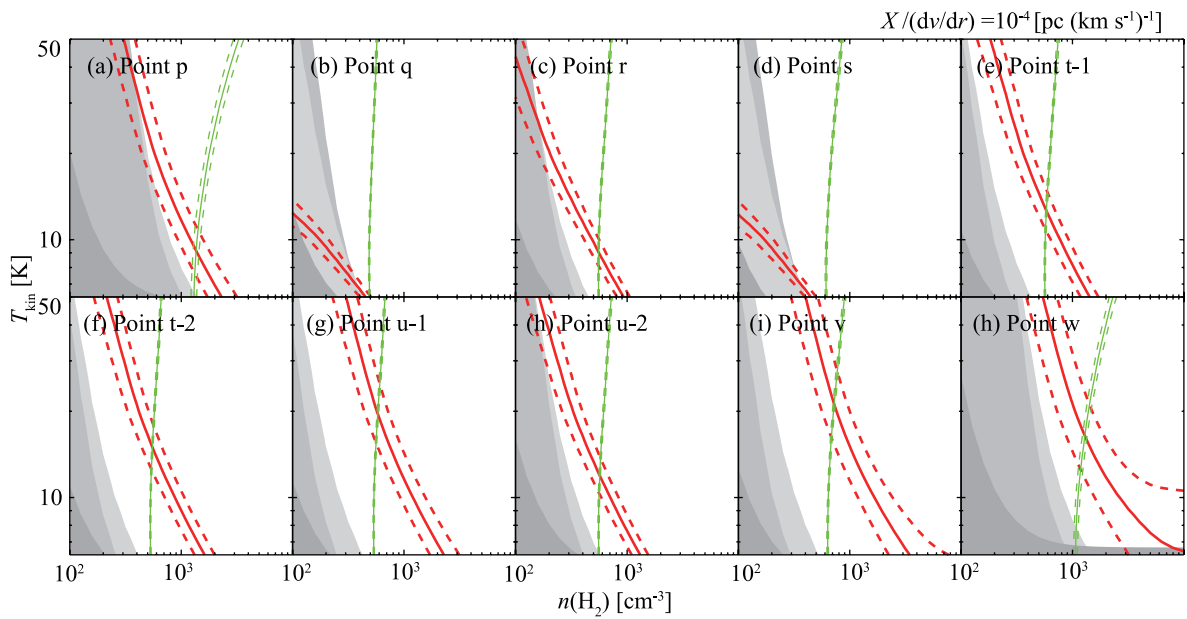}
\caption{
Number density and kinetic temperature curves obtained by LVG analyses.
The thick and thin (red and green in a color version) lines are the observed ratios of $^{13}$CO($J$=2--1)/$^{13}$CO($J$=1--0) and $^{13}$CO($J$=2--1)/$^{12}$CO($J$=2--1), respectively.
The dashed lines indicate errors from calibration errors of each line, of 10\% for $^{12}$CO($J$=2--1) and $^{13}$CO($J$=1--0) and 20\% for $^{13}$CO($J$=2--1).
The gray zones indicate areas where the filling factors become above 1 (dark: $^{12}$CO $J$=2--1, middle: $^{13}$CO $J$=1--0, and light: $^{13}$CO $J$=2--1; see Section \ref{subsubsec:lvg}).
\label{lvg}}
\end{figure*}

We chose eight local peaks p -- w of $^{12}$CO($J$=1--0) for the $^{13}$CO($J$=2--1) pointed observations. Their positions and line profiles are shown in Figures \ref{ratio_2-1_1-0c} and \ref{spectra}, respectively.
We used averaged intensity ratios around the barycenter velocity with a 1\,km\,s$^{-1}$ window in the LVG analysis. We chose two combinations of the ratios among $^{12}$CO($J$=2--1), $^{13}$CO($J$=1--0) and $^{13}$CO($J$=2--1).
We did not use the $^{12}$CO($J$=1--0) line since it is generally optically thick and may sample only the nearside lower density gas \citep[e.g.,][]{2010PASJ...62...51M}.
The averaged intensities and barycentric velocities are listed in Columns (6) -- (9) of Table \ref{tlvgprop}.
For the points t and u, where spectra show double components, we estimate $T_{\rm kin}$ and $n({\rm H}_2)$ for each component.
$X/(dv/dr)$ is set to $10^{-4}$\,pc\,(km\,s$^{-1}$)$^{-1}$ with the canonical [CO]/[H$_2$] abundance of $10^{-4}$ \citep{1982ApJ...262..590F}.
The results for $T_{\rm kin}$ and $n({\rm H}_2)$ are shown in Figure \ref{lvg}.
The curves represent each observational line intensity ratio.
We also estimated a beam filling factor through dividing the observed line intensity by the calculated line intensity for the three transitions.
If this factor becomes greater than 1, the parameter will not be realized (gray areas in Figure \ref{lvg}).

The density is estimated to be less than 10$^3$\,cm$^{-3}$ except for the points p, v, and w.
At the peak position of the {\it arc} cloud and J3, i.e. the points p and w, the density increases to $\sim$10$^{3.1}$\,cm$^{-3}$.
The {\it arc} cloud and J1, i.e., points p -- s, have low temperature below $\sim$10\,K, whereas the other positions in the {\it jet} cloud, especially v and w, have higher temperatures of $\sim$20\,K in a range of 12 -- 28\,K.

\subsection{Star Formation in the {\it Jet} and {\it Arc} Clouds}\label{subsec:starformation}

\begin{figure}
\includegraphics{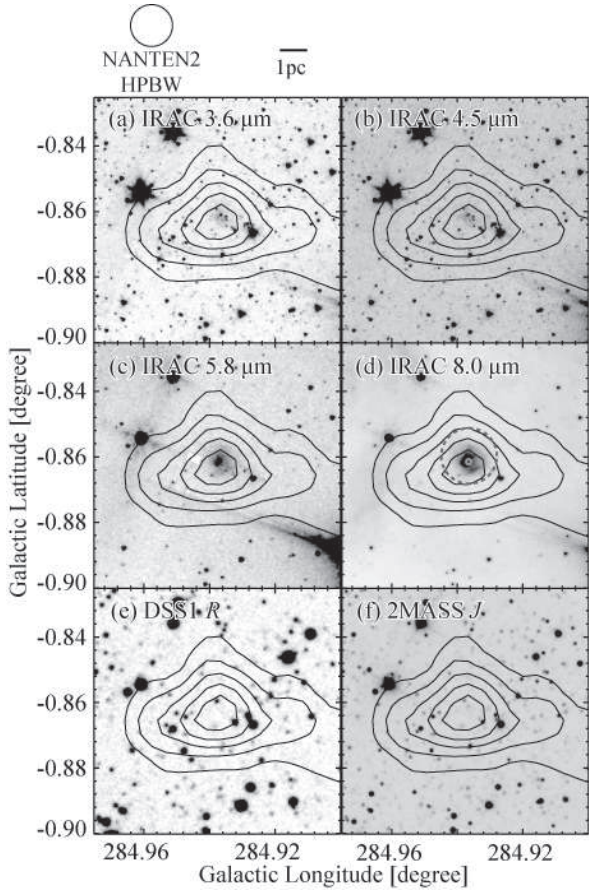}
\caption{
Near -- middle infrared ({\it Spitzer}/IRAC 3.6, 4.5, 5.8, 8.0\,{\micron} and 2MASS $J$ band) and visible (DSS1 $R$ band) images toward J3a.
The contours are the $^{13}$CO($J$=1--0) integrated intensity drawn every 1.0\,K\,km\,s$^{-1}$ (=3$\sigma$) from 1.4\,K\,km\,s$^{-1}$ (=4$\sigma$).
The small circle and large dashed-line circle show positions of point sources in 9.0\,{\micron} from {\it AKARI} IRC PSC and in 160\,{\micron} from {\it AKARI} FIS BSC.
The diameters of the circles indicate each angular resolution (FWHM) of 5{\farcs}5 in 9.0\,{\micron}  \citep{2007PASJ...59S.401O} and 61{\arcsec} for in 160\,{\micron} \citep{2007PASJ...59S.389K}, respectively. 
\label{star_formation}}
\end{figure}

We have searched for young objects in the {\it IRAS} point source catalog and the {\it Siptzer}/IRAC data from Spitzer Heritage Archive.
As a result, we find one possible candidate for a young star toward J3a as shown in Figure \ref{star_formation}.
The source is detected significantly toward the $^{13}$CO ($J$=1--0) peak as a nebulous extended source at 5.8\,{\micron} and 8.0\,{\micron}.
This source is possibly identical with sources detected at 9\,{\micron} by the {\it AKARI} IRC Point Source Catalogue (IRC PSC) \citep{AKARI-IRC-PSC} and 160\,{\micron} by {\it AKARI} FIS Bright Source Catalogue (FIS BSC) \citep{AKARI-FIS-BSC}, whereas no counterpart can be confirmed in DSS1 Red band, 2MASS $J$ band (Figures \ref{star_formation} e and f) and radio continuum at 843\,MHz (Figure \ref{fig1}b).
The flux densities of the {\it AKARI} sources are summarized in Table \ref{akari}.
By the lack of infrared data at the other wavelengths, especially far-infrared, a detailed analysis of the evolutionary stage (e.g., Class 0--III) and estimation of the mass of the young star remain as the future work.
It takes $\sim$Myr for such usual low-mass star formation, as is consistent with the crossing time of the {\it jet} cloud of a few Myr.

Figure \ref{fig1}b shows a radio continuum image at 843\,MHz from the 2nd epoch Molonglo Galactic Plane Survey (MGPS-2) by the Molonglo Observatory Synthesis Telescope (MOST) \citep{2007MNRAS.382..382M} with the integrated intensity of the {\it jet} and {\it arc} clouds in $^{12}$CO ($J$=1--0) overlaid.
The radio emission is dominated by the thermal free-free emission of the \ion{H}{2} region RCW 49.
The \ion{H}{2} region shows a ridge extending toward the southeast nearly along the {\it jet} cloud axis, while its connection with the {\it jet} cloud is not clear.
We find no significant radio source except for a few unresolved compact sources in the direction of the {\it jet} clouds at $(l, b)=$(284\fdg 53, $-0\fdg 68$) and $(l, b)=$(284\fdg 68, $-0\fdg 74$).
However, no counterparts of these radio sources are seen in the infrared data such as the {\it Siptzer}/IRAC, which show the hot dust heated by nearby \ion{H}{2} regions.
In addition, $R_{2-1/1-0}$ in Figure \ref{ratio_2-1_1-0} is not high toward the position of the radio sources, and we find no significant velocity variations around these two radio sources at $S \sim 0\fdg45$ and $0\fdg56$ as seen in Figures \ref{stmap_12co21} and \ref{stmap_12co10}.
Therefore, it is likely that the radio sources are not compact \ion{H}{2} regions associated with the {\it jet} cloud and may be extragalactic.
To summarize, the {\it jet} and {\it arc} clouds as a whole are not active in star formation.

\subsection{Comparison with the TeV $\gamma$-Rays}\label{subsec:comparison_gamma}

\begin{figure*}
\includegraphics[height=11cm]{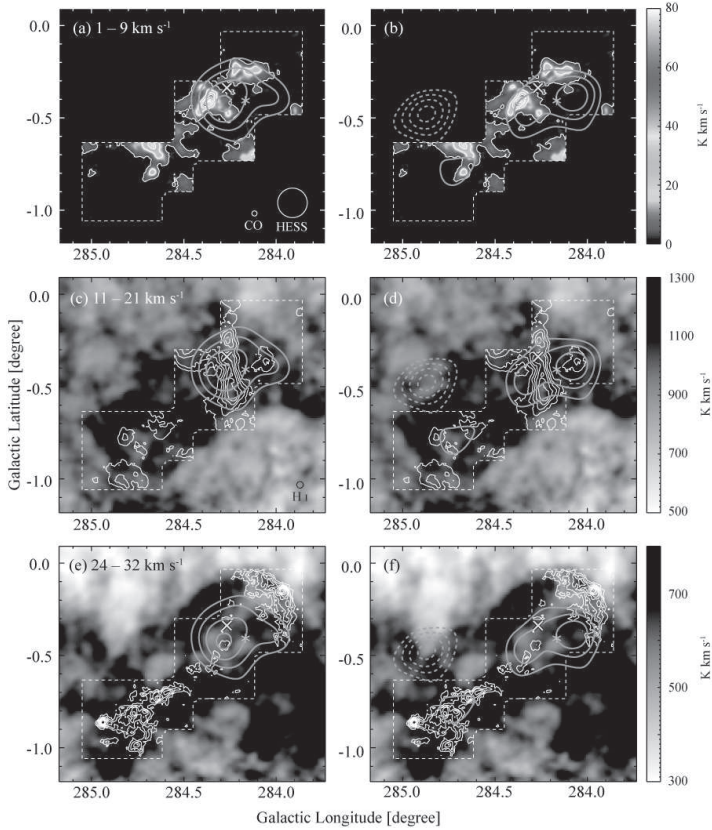}
\caption{ 
Comparison between TeV $\gamma$-ray distribution and ISM gas.
The thick (orange in a color version) contours are the TeV $\gamma$-ray smoothed excess images \citep[as same as Figure 2 in][]{2011A&A...525A..46H}.
The left side panels (a, c, and e) are at the low energy band (0.7 -- 2.5\,TeV), and the right side panels (b, d, and f) are at the high energy band (above 2.5\,TeV).
The contours are drawn every 10 excess arcmin$^{-2}$ from 40 excess arcmin$^{-2}$ for the low energy band, and every 5 excess  arcmin$^{-2}$ from 20 excess arcmin$^{-2}$ for the high energy band.
The dashed contours show HESS J1026$-$582.
(a) and (b) The images and thin contours show the integrated intensities of $^{12}$CO($J$=2--1) in the velocity range of the GMCs at 4\,km\,s$^{-1}$.
The contours are drawn every 14\,K\,km\,s$^{-1}$ (=40$\sigma$) from 2.8\,K\,km\,s$^{-1}$ (=8$\sigma$).
(c) and (d) The gray scale shows the integrated intensities of the \ion{H}{1} 21\,cm line in the velocity range of the GMCs at 16\,km\,s$^{-1}$.
The thin contours are the integrated intensities of $^{12}$CO($J$=2--1) drawn every 5.8\,K\,km\,s$^{-1}$ (=15$\sigma$) from 3.1\,K\,km\,s$^{-1}$ (=8$\sigma$).
(e) and (f) The gray scale shows the integrated intensities of the \ion{H}{1} 21\,cm line in the velocity range of the {\it jet} and {\it arc} clouds.
The thin contours are the integrated intensities of $^{12}$CO($J$=2--1) drawn every 1.5\,K\,km\,s$^{-1}$ (=4$\sigma$).
The symbols are the same as in Figure \ref{fig1}.
\label{origin}}
\end{figure*}

In order to see if the $\gamma$-rays show spatial correspondence with the molecular gas, we show a comparison of the TeV $\gamma$-ray distribution of HESS J1023$-$575 \citep{2011A&A...525A..46H} with the CO at three velocity intervals in Figure \ref{origin}.
We show the two energy ranges of 0.7 -- 2.5\,TeV (low-energy band) and of above 2.5\,TeV (high-energy band).
A source HESS J1026$-$582 shown by the dashed line is supposed to be a PWN powered by a pulsar PSR J1028$-$5819 at a distance 2.3\,kpc and is not associated with Wd 2 \citep{2011A&A...525A..46H}.

We find that the major parts of the two GMCs at 1 -- 21\,km\,s$^{-1}$ \citep{2009ApJ...696L.115F} are within the lowest contour of the $\gamma$-rays (Figures \ref{origin} a -- d).
Considering the relatively coarse resolution of H.E.S.S.\ $\sim 0\fdg 07$, the two GMCs may be possible candidates for the $\gamma$-ray production via the hadronic process \citep{2009ApJ...707L.179F, 2011A&A...525A..46H}.
On the other hand, the {\it jet} and {\it arc} clouds are located around HESS J1023$-$575, showing an anti-correlation with the $\gamma$-ray distribution, whereas there is relatively bright \ion{H}{1} gas toward the $\gamma$-rays (Figures \ref{origin} e and f).
The target protons may be mostly atomic, if the $\gamma$-rays are of the hadronic origin from the {\it jet} and {\it arc} clouds region.
The distribution at the high-energy TeV $\gamma$-ray band shows a hint of elongation along the {\it jet} cloud direction, suggesting that the $\gamma$-ray source may be related with the {\it jet} cloud.
In the high energy band, another faint source is detected about 4$\sigma$ but less than 5$\sigma$ toward the {\it jet} cloud at $(l, b) = (284\fdg 7, -0\fdg 8)$, possibly suggesting association with the {\it jet} cloud.
Toward this new $\gamma$-ray source only a radio source at $(l, b)=$(284\fdg 68,$-0\fdg 74$) in Figure \ref{fig1}b exists within the 4$\sigma$ contour.
However, there is no identification of the radio source in \citet{2010A&A...518A..10V} and  \citet{2005AJ....129.1993M} so far.
The radio source may also be a possible candidate of the new $\gamma$-ray source counterpart, but this cannot be confirmed.
We discuss further on the $\gamma$-ray origin toward the Wd 2 region in Section \ref{subsec:origin_gamma}.


\section{DISCUSSION}\label{sec:discussion}

\subsection{The {\it Jet} and {\it Arc} Clouds Associated with HESS J1023$-$575}

Most of the individual {\it jet} and {\it arc} clouds have density, temperature and linewidth similar to the other Galactic molecular clouds like the Taurus dark cloud.
The highly filamentary and elongated nature of the {\it jet} cloud having 100\,pc length and several pc width may, however, be fairly exceptional among the Galactic clouds; the typical filamentary clouds have shorter projected lengths of 10 -- 50\,pc (e.g., Taurus: \citealt{1995ApJ...445L.161M}; Chamaeleon: \citealt{2001PASJ...53.1071M}; L1333: \citealt{1998AJ....115..274O}).
It is also notable that the {\it arc} cloud having $10^4$\,$M_\sun$ shows an outstanding crescent shape; we have not yet been informed on such a well-defined arc-like cloud to date in the literature.
If the blue-shifted features J2 -- J4 and the red-shifted features J3 -- J5 are physically connected, part of the {\it jet} cloud has a velocity span of $\sim$10\,km\,s$^{-1}$.
These broad features may not be gravitationally bound by the self-gravity for the cloud mass and size of $\sim10^4$\,$M_\sun$ and $\sim$10\,pc, respectively (Columns 3 and 4 in Table \ref{phys_param}).
We shall discuss the origin of the {\it jet} and {\it arc} clouds later into more detail.

The molecular clouds toward Wd 2 consist of two groups of different velocity ranges, $-10$ -- 20\,km\,s$^{-1}$ and 20 -- 30\,km\,s$^{-1}$.
The former group includes the two parent GMCs at 4\,km\,s$^{-1}$ and 16\,km\,s$^{-1}$ which show tight correlation with Wd 2 and RCW 49 as verified by the high line intensity ratio of $^{12}$CO ($J$=2--1)/($J$=1--0) around 1.0, corresponding to kinetic temperature above $\sim$30\,K \citep{2010ApJ...709..975O}.
The latter group consisting of the {\it jet} and {\it arc} clouds shows no enhanced line ratio of $^{12}$CO ($J$=2--1)/($J$=1--0) or higher temperature toward RCW 49 (points q and r in Figure \ref{ratio_2-1_1-0c}).
We therefore infer that the {\it jet} and {\it arc} clouds are not close enough to be heated by RCW 49 or Wd 2 and this geometrical separation is consistent with the velocity difference of the two groups by $\sim$20\,km\,s$^{-1}$ and with a kpc-scale separation between them. 

By comparison with the TeV $\gamma$-ray distributions in Figure \ref{origin}, we suggest that the {\it jet} and {\it arc} clouds are physically associated with HESS J1023$-$575 and are located at a distance of 7.5\,kpc, larger than that of Wd 2 of 5.4\,kpc.
A consequence of this location is that the scenarios for $\gamma$-rays production due to the stellar winds/blister energized by the stellar cluster may not be appropriate. 

In the followings, we shall discuss on 1) the formation of the {\it jet} cloud, 2) the formation of the {\it arc} cloud, and 3) the possible origin for the $\gamma$-ray source. 

\subsection{Formation of the {\it Jet} Cloud}\label{subsec:formation_jet}

Paper I presented a hypothesis about the formation mechanism of the {\it jet} cloud as follows: first, a high-energy jet driven by an anisotropic SN explosion or by a microquasar is injected into the ambient atomic or molecular gas.
The gas is then compressed cylindrically by the shock fronts due to the high-energy jet.
Eventually, molecular clouds with a linear shape were formed in the compressed gas.

Asahina et al.\ (2013) carried out magneto-hydro-dynamical (MHD) numerical simulations of the interaction between a high-energy jet and ambient ISM.
As the initial conditions, they assume \ion{H}{1} gas in the Cold Neutral Medium (CNM) phase with temperature of $\sim$200\,K and density of $\sim$10\,cm$^{-3}$.
The velocity and radius of the high-energy jet were assumed to be 200\,km\,s$^{-1}$ and 1\,pc, which aims at modeling the high-energy jet slowed down in the interaction with the ISM, or, the high-energy jet launched by a rotating disk much larger than the size of the compact object.
Such a jet is actually consistent with SS~433, the microquasar having the X-ray jet \citep{1996A&A...312..306B} extending by $\sim$50\,pc from the central object at a distance of $\sim$3\,kpc (\citealt{1998AJ....116.1842D}; a different distance of 5.5\,kpc is estimated by \citealt{1981ApJ...246L.141H,2004ApJ...616L.159B}).
When the high-energy jet is injected, the \ion{H}{1} gas is compressed and heated up, where this heating will be on a scale much smaller than pc.
In the calculations, a cooling function defined in \citet{2006ApJ...652.1331I} is adopted, whereas the self-gravity is not included.
By the cooling effect, the compressed gas finally becomes cool and dense with temperature of $\sim 50$\,K and density of $\sim 50$\,cm$^{-3}$ due to the thermal instability.
It is possible that such cool and dense gas can be transformed by self-gravity into the molecular clouds which trace the trajectory of the high-energy jet.
The column density of the \ion{H}{1} gas is actually enhanced toward the {\it jet} cloud as shown in Figure \ref{jet} (a).
Enhancement of the \ion{H}{1} gas column density along the {\it jet} cloud axis is also seen in SS 433 and in another candidate for the molecular jet at $(l, b)=$(348\fdg 5, $-2\degr$ -- $+2\degr$) \citep{2008PASJ...60..715Y}.
Asahina et al.\ (2013) have shown that the compressed gas has an expansion velocity of $\sim$2\,km\,s$^{-1}$ perpendicular to the propagation axis of the high-energy jet.
This is consistent with the full linewidths of the individual {\it jet} clouds of $\sim2$ -- 5\,km\,s$^{-1}$ (Column 10 in Table \ref{obs_para12co10}).
Although the anisotropic jet-like SN explosion is suggested in Paper I, there is no numerical simulation on such a process.
We shall not discuss this alternative further in the present work.
The numerical simulations by Asahina et al.\ (2013) may also be applicable to the anistropic jet-like SN explosion if the physical parameters such as the jet speed and the shock strength are similar to those in Asahina et al.\ (2013).

The interaction may cause heating of the gas, while the cooling by molecule lines is fairly rapid like $10^4$\,yrs or less for density around $10^3$\,cm$^{-3}$ or higher as derived in the present LVG analysis shown in Figure \ref{lvg} and Table \ref{tlvgprop} \citep[e.g.,][]{1978ApJ...222..881G}.
In Section \ref{subsubsec:intensity_ratio} the distribution of the line intensity ratio $^{12}$CO ($J$=2--1)/($J$=1--0) is shown in Figure \ref{ratio_2-1_1-0}, which indicates that the kinetic temperature may become higher in a few places in the {\it jet} cloud, i.e., toward J3a, J3b, J4 and J5. J3a and J3b show line ratios of 0.6 -- 0.8 implying temperature of $\sim$20\,K, significantly higher than the usual temperature of the molecular gas, 10\,K, with no extra local heating.
We also note that parts of J4 and J5 show somewhat higher line intensity ratios above 0.7, and they may suggest some extra shock heating due to the high-energy jet.
A possibility is that J4 and J5 accelerated by the high-energy jet are still interacting with other molecular gas and have been causing a shock heating until now.
J3a has a sign of recent star formation as shown in Section \ref{subsec:starformation} and it may be an alternative that this young star is radiatively heating J3a, while the star with no thermal radio continuum source (Figure \ref{fig1}b) may not be luminous enough to explain the temperature excess.

The winding shape of the {\it jet} cloud in a few places is intriguing (Sections \ref{subsubsec:global_distribution} and \ref{subsubsec:details_jet}) and may give some hints on the details of the interaction of the high-energy jet.
For instance, the winding shape may trace  MHD instability of a helical shape in the propagating magnetized jet. Such helical patterns are in fact observed in MHD numerical simulations of the high-energy jet \citep[e.g.,][]{2001NewA....6...61N}.
It is also interesting to see the time variation of the high-energy jet; a question is if the jet is impulsive or intermittent over a time scale of Myr.
It may be difficult to explain the winding shape of the remnant molecular gas by a continuous helical flow of the high-energy jet material, which may destroy the helical pattern even if it is once formed.
Because of this, impulsive or intermittent jet may be favorable.

\subsection{Formation of the {\it Arc} Cloud}\label{subsec:formation_arc}

We shall then discuss the formation of the {\it arc} cloud.
The present work has shown two new detailed aspects of the {\it arc} cloud.
One is the western {\it jet} cloud identified toward the {\it arc} cloud, which seems to be terminated in the middle of the {\it arc} cloud, and the other is that the {\it arc} cloud shows symmetry with respect to the {\it jet} cloud axis.
The {\it arc} cloud is also part of the \ion{H}{1} shell as suggested in Paper I (see Figure 3 of Paper I).

An obvious explanation of the formation of the {\it arc} cloud is an SN explosion.
In order to test this possibility we calculate the radius and mass of a shell formed due to an SN explosion by using equations in Appendix \ref{sec:snrevolution}.
We shall assume a uniform ambient density, $10^{51}$\,erg for the total mechanical energy of the SN explosion, and $10^4$\,km\,s$^{-1}$ for the initial velocity of the shock, and calculate the parameters when the shock velocity decreases to the sound speed of typical temperature $\sim$100\,K in the CNM, i.e. $\sim$1\,km\,s$^{-1}$.
According the calculations, when the ambient density is $\sim$1\,cm$^{-3}$, the radius reaches 45\,pc, which is consistent with that of the {\it arc} cloud at the distance of 7.5\,kpc, for an age of $5 \times 10^4$\,yr.
But the mass of the shell is only $ 9.1 \times 10^3$\,$M_\sun$, which is much smaller than the observed total mass of the {\it arc} cloud and \ion{H}{1} shell $\sim2 \times 10^5$\,$M_\sun$.
To explain the mass of the {\it arc} cloud and \ion{H}{1} shell, the ambient density of 15\,cm$^{-3}$ is required, but, if so, the radius of an SNR becomes smaller as 34\,pc for an age of $1 \times 10^5$\,yr.
If the mechanical energy of the SN explosion is higher by a factor of 5, i.e., $5 \times 10^{51}$\,erg, both of the radius and mass of the {\it arc} cloud and \ion{H}{1} shell can be explained even if the ambient density is 15\,cm$^{-3}$.
However, we should note that the calculations include many assumptions such as the uniform density, and the real situations may be more complicated.
In addition, we have another problem of the column density contrast between the \ion{H}{1} shell and the CO arc of about 1:10.
It is questionable if the shell can expand to a similar radius for such density contrast.
Therefore, we have to note that there are difficulties on the SN explosion to form the {\it arc} cloud.

As an alternative possibility of the formation mechanism of the {\it arc} cloud, we shall consider the interaction of the high-energy jet.
The shape of the {\it arc} cloud is symmetric about the {\it jet} cloud axis, and the western {\it jet} cloud is elongated along the {\it jet} cloud axis.
These properties suggest that the high-energy jet is tightly related to the formation of the {\it arc} cloud.
Asahina et al.\ (2013) shows that the high-energy jet forms a bow shock in the adiabatic case.
This suggests that, if the cooling is not dominant, the interaction may form a bow-shock like {\it arc} cloud at the edge of the western {\it jet} cloud.
A ring-like radio structure similar to the {\it arc} cloud is observed around the microquasar Cyg-X1 \citep{2005Natur.436..819G} and it is believed that the interaction between the high-energy jet and surrounding ISM created the structure.
This radio emission originates from the bremsstrahlung from hot plasma, and some atomic lines such as H$\alpha$ and \ion{O}{3} are also detected in the radio ring \citep{2007MNRAS.376.1341R}.
If such hot gas cools down, molecular gas like the {\it arc} cloud may be formed. It is desirable that a further study is devoted to pursue the formation of the bow shock driven by the high-energy jet via numerical simulations in order to have a better insight into the bow shock formation.
It is puzzling that the eastern and western {\it jet} clouds are fairly asymmetric in the sense that the eastern {\it jet} cloud is extended by a factor of 2.5 than the western {\it jet} cloud.
The {\it arc} cloud suggests that the ISM in the western region is more massive than in the eastern region of Wd2 and may offer a possible explanation on the asymmetry.

\subsection{Origin of the $\gamma$-Rays}\label{subsec:origin_gamma}

We shall discuss the origin of the $\gamma$-rays of HESS J1023$-$575 in the present framework based on the {\it jet} and {\it arc} clouds.
Since the $\gamma$-rays are distributed inside the \ion{H}{1} shell, where there is no CO, in Figure \ref{origin}, the \ion{H}{1} gas is considered as the targets for CR protons, if the $\gamma$-rays are of the hadronic origin.
The basic timescales of the hadronic process are the cooling timescale and the diffusion timescale.
Density of the target protons in a sphere with radius of 0{\fdg}18 centered at the central position of HESS J1023$-$575 is derived to be $n_\mathrm{inner}$=12\,cm$^{-3}$ at 7.5\,kpc.
Then, the cooling time scale due to the pp-interaction is given as follows \citep{2009MNRAS.396.1629G},
\begin{equation}
t_{\rm pp}=5 \times 10^6\, {\rm yr} \left( \frac{n_{\rm inner}}{12\, {\rm cm}^{-3}} \right)^{-1}.
\label{eqn:tpp}
\end{equation}
Also, since the radius of the $\gamma$-ray extent (0\fdg 18) becomes $R_\mathrm{inner}$=24\,pc at 7.5\,kpc, the diffusion time scale of CR protons is given as follows \citep{2009MNRAS.396.1629G},
\begin{equation}
t_{\rm diff}=2 \times 10^4\, {\rm yr} \left( \frac{\chi}{0.01} \right)^{-1} \left( \frac{R_{\rm inner}}{24\, {\rm pc}} \right)^2 \left( \frac{E_{\rm p}}{10\, {\rm TeV}} \right)^{-0.5} \left( \frac{B}{10\, {\mu \rm G}} \right)^{0.5},
\label{eqn:tdiff}
\end{equation}
where $E_\mathrm{p}$=10\,TeV is the CR proton energy, the magnetic field strength $B$ is assumed to be 10\,$\mu$G and $\chi$ means the deviation from typical diffusion coefficient in the Galaxy.
If an SN explosion occurred in the past, magnetic turbulence will be enhanced and yield low $\chi$ value such as 0.01.
Equations (\ref{eqn:tpp}) and (\ref{eqn:tdiff}) show that the diffusion timescale is much shorter than the other timescales although the cooling timescale is consistent with the crossing timescale of the {\it jet} and {\it arc} clouds.
Therefore, the particle acceleration should occurs within $\sim 10^{4}$\,yr in the hadronic scenario.

Most known TeV $\gamma$-ray sources in the Galaxy are either SNRs or PWNe.
The bright TeV $\gamma$-ray SNRs are all young SNRs of 1000 -- 2000\,yrs (RX J1713.7$-$3946: \citealt{2006A&A...449..223A, 2007A&A...464..235A}; RX J0852.0$-$4622: \citealt{2007ApJ...661..236A}; RCW86: \citealt{2009ApJ...692.1500A} and HESS J1731$-$347: \citealt{2011A&A...531A..81H}) and the TeV $\gamma$-ray PWNe are younger than $\sim10^5$\,yr as estimated by the spin-down rate \citep{2010AIPC.1248...25K}.
On the other hand, toward Wd 2, no SNRs are known, and conclusive evidence of a PWN has not been found.
Even if HESS J1023$-$575 is either an SNR or PWN associated with the {\it jet} and {\it arc} clouds, lifetime of the relativistic particles accelerated at the SNR or PWN are much shorter than the age of the {\it jet} and {\it arc} clouds of $\sim$Myr suggested by the crossing timescale and the existence of star formation.
If an SN which is a progenitor of the SNR or PWN created the {\it jet} and {\it arc} clouds in $\sim$Myr ago, the relativistic particles from the SNR or PWN must have already cooled or escaped in the first $\sim10^4$\,yr, and cannot emit the TeV $\gamma$-rays at present.
The same reasoning will be applied to the hypothesis, an anisotropic SN explosion (Sections \ref{subsec:formation_jet} and \ref{subsec:formation_arc}).
Therefore, an SNR or a PWN is not likely as the origin for HESS J1023$-$575, if the {\it jet} and {\it arc} clouds and HESS J1023$-$575 are due to a single object.

The only remaining possible candidate for the $\gamma$-ray origin is the CRs accelerated in microquasar jet interacting with the ambient ISM.
Particle acceleration of the termination shock of microquasar jet interacting with the ISM has been theoretically studied and it is shown that such interaction can produce CRs \citep[e.g.,][]{2005A&A...432..609B}.
The microquasar may have a potential to be active over $\sim$Myr, even intermittently, which favors the molecular cloud formation.
Figures \ref{origin} (e) and (f) indicate that there exists relatively dense \ion{H}{1} gas beyond the TeV $\gamma$-ray extent, suggesting that the spatial extent of the CR protons should be similar to that of the TeV $\gamma$-rays.
The CRs may be confined by the dense \ion{H}{1} shell and the {\it arc} cloud.
Total energy of the CR protons estimated from the TeV $\gamma$-ray luminosity $L_\mathrm{\gamma}$(1--10\,TeV) is given as follows;
\begin{equation}
W_\mathrm{p} \left( 10\mbox{--}100\, {\rm TeV} \right) \sim 7 \times 10^{48}\, {\rm erg} \left( \frac{t_{\rm pp}}{5 \times 10^6\, {\rm yr}} \right) \left[ \frac{L_\gamma \left( 1\mbox{--}10\, {\rm TeV} \right) }{4.5 \times 10^{34}\, {\rm erg\, s}^{-1}} \right].
\end{equation}
Here, we use the $\gamma$-ray luminosity $4.5 \times 10^{34}$\,erg\,s$^{-1}$ derived at 7.5\,kpc. 
If this energy is released during the diffusion time of $\sim$10$^{4}$\,yr, averaged power injected to the CR protons becomes $\sim2 \times 10^{37}$\,erg\,s$^{-1}$ which is consistent with the theoretical work \citep{2005A&A...432..609B}.

The total radio fluxes of the known microquasors (e.g., SS~433, LS~I~+61~303, LS~5039 and Cyg~X-3) located within several kpc are measured to be 867\,mJy, 42\,mJy, and 87\,mJy at 1.4\,GHz, respectively \citep{2002A&A...394..193P}.
On the other hand, the observed flux toward RCW49 is larger than 1\,Jy\,beam$^{-1}$ at 843\,MHz as seen in Figure \ref{fig1}.
It is thus hard to distinguish a microquasar from the strong and complicated radio emission,  if it not exceptionally bright like SS 433.

We should note that the physical characteristics of HESS J1023$-$575 are different from those of the other microquasars in the Galaxy. The $\gamma$-rays toward microquasars have been discovered at Cyg-X1 \citep{2007ApJ...665L..51A, 2010ApJ...712L..10S} and Cyg-X3 \citep{2009Sci...326.1512F}.
They are compact (unresolved) and have a time modulation or flares. The $\gamma$-ray binaries, such as LS 5039 \citep{2006A&A...460..743A, 2009ApJ...706L..56A} and LS I+61{\degr}303 \citep{2006Sci...312.1771A, 2008ApJ...679.1427A, 2009ApJ...701L.123A}, which are thought to be candidates of microqusars, also have compact features and time modulations.
We infer that the $\gamma$-ray radiation mechanisms are different  between HESS J1023$-$575 and the other microquasars in physical conditions such as density of the ambient matter. 

As shown in Section \ref{subsec:comparison_gamma}, the faint $\gamma$-ray emission was discovered toward the {\it jet} cloud J3 at $(l,b)= (284\fdg 7, -0\fdg 8)$ by the new H.E.S.S.\ observations.
A counterpart such as a pulsar or blazar toward the source has not been found so far.
Since this $\gamma$-ray spot is $\sim 0\fdg 6$ ($\sim$60\,pc at 5.4\,kpc and $\sim$30\,pc at 2.4\,kpc) away from Wd 2 and PSR J1023$-$5746, the cluster or the pulsar cannot be responsible for the $\gamma$-ray emission.
If the $\gamma$-ray emission toward the {\it jet} cloud is of the hadronic origin by the reaccelerated protons, the cooling timescale and diffusion timescale are estimated to be as follows;
\begin{equation}
t_{\rm pp} \sim 6 \times 10^4\, {\rm yr} \left( \frac{n_{\rm J3}}{1000\, {\rm cm}^{-3}} \right)^{-1} 
\end{equation}
\begin{equation}
t_{\rm diff} \sim 7 \times 10^3\, {\rm yr} \left( \frac{\chi}{0.01} \right)^{-1} \left( \frac{R_{\rm J3}}{16\, {\rm pc}} \right)^2 \left( \frac{E_{\rm p}}{10\, {\rm TeV}} \right)^{-0.5} \left( \frac{B}{10\, {\mu \rm G}} \right)^{0.5}
\end{equation}
at 7.5\,kpc.
Here, $n_{\rm J3}$ is number density of the J3 cloud obtained by the LVG analysis in Table \ref{akari}, and $R_{\rm J3}$ is the radius of J3 in Table \ref{phys_param}.
By assuming that the luminosity of the $\gamma$-ray source is an order of magnitude less than that of HESS J1023$-$575, the total energy of the reaccelerated protons is estimated as 
\begin{equation}
W_\mathrm{p} \left( 10\mbox{--}100\, {\rm TeV} \right) \sim 2 \times 10^{45}\, {\rm erg} \left( \frac{t_{\rm pp}}{6 \times 10^4\, {\rm yr}} \right) \left[ \frac{L_\gamma \left( 1\mbox{--}10\, {\rm TeV} \right) }{10^{33}\, {\rm erg\,s}^{-1}} \right]
\end{equation}
If this total energy of the reaccelerated protons is released for the diffusion time of $7 \times 10^3$\,yr, averaged power injected to the relativistic protons is derived to be $\sim 8 \times 10^{33}$\,erg\,s$^{-1}$.
The kinetic power of a microquasar jet like SS 433 is more than enough to explain such power.


\section{CONCLUSION}\label{sec:conclusion}
We summarize the present work as follows;
we presented detailed CO ($J$=2--1 and 1--0) distributions of the {\it jet} and {\it arc} clouds which were discovered toward Wd 2 by Paper I.
The {\it jet} cloud consists of two components, the eastern {\it jet} cloud and the western {\it jet} cloud, well aligned on an axis passing near the peak of the TeV $\gamma$-ray source HESS J1023$-$575.
The total length of the {\it jet} cloud is $\sim$140\,pc in the sky.
The eastern {\it jet} cloud shows unique winding distributions and a bifurcation.
The western {\it jet} cloud has been resolved toward the {\it arc} cloud by the present work.
The {\it arc} cloud shows a crescent shape whose velocity distribution does not indicate a sign of expansion.
The {\it arc} cloud seems to constitute part of the \ion{H}{1} shell having a size of 30 -- 40\,pc, which appears to surround HESS J1023$-$575.
The {\it jet} and {\it arc} clouds show strong correspondence with the TeV $\gamma$-rays but their correlation with Wd 2 is not clear.
We therefore associate the {\it jet} and {\it arc} clouds with HESS J1023$-$575 but not necessarily with Wd 2.
Rather, the kinematic distance of the {\it jet} and {\it arc} clouds is estimated to be 7.5\,kpc, whereas that of Wd 2 is 5.4\,kpc, suggesting that the {\it jet} and {\it arc} clouds are significantly separated from Wd 2.
If this is the case, the stellar wind collision by Wd 2 is not a viable scenario in the $\gamma$-ray production toward Wd 2.

 The {\it jet} cloud shows a sign of star formation on its edge but the {\it arc} cloud shows no sign for star formation.
The temperature of the {\it jet} and {\it arc} clouds is estimated toward eight selected positions by the LVG analysis of the three transitions, $^{12}$CO ($J$=2--1) and $^{13}$CO ($J$=2--1 and $J$=1--0).
The results are consistent with kinetic temperature around 10\,K and molecular hydrogen density around $10^3$\,cm$^{-3}$, whereas, toward a few positions in the {\it jet} cloud, we find that temperature is as high as 20\,K, possibly suggesting some additional heating like that due to the shock interaction. 
 
 The TeV $\gamma$-rays toward Wd 2 show a strong correspondence with the {\it jet} and {\it arc} clouds instead of Wd 2 and RCW 49.
This association favors a distance of 7.5\,kpc for the $\gamma$-ray source, the same as the {\it jet} and {\it arc} clouds.
If an event formed both of the {\it jet} and {\it arc} clouds and HESS J1023$-$575, SNRs and PWNe are not able to explain the present $\gamma$-rays emission because of short lifetime of the accelerated particles compared with the age of the clouds.
We discuss two possible hypotheses on the origin of the {\it jet} and {\it arc} clouds and the TeV $\gamma$-ray emission,  1) a microquasar jet or 2) an anisotropic SN explosion with a microquasar jet driven by the stellar remnant.
For these scenarios it is necessary to assume that the microquasar is active over $10^6$\,yrs as the origin of the current TeV $\gamma$-rays, whereas such a microquasar has not yet been observed elsewhere in the Galaxy.
HESS J1023$-$575 remains as one of the most enigmatic TeV $\gamma$-ray sources.

\acknowledgments
NANTEN2 is an international collaboration of 10 universities: Nagoya University, Osaka Prefecture University, University of Cologne, University of Bonn, Seoul National University, University of Chile, University of New SouthWales, Macquarie University, University of Sydney, and University of ETH Zurich.
The Mopra radio telescope is part of the Australia Telescope National Facility which is funded by the Commonwealth of Australia for operation as a National Facility managed by CSIRO. The University of New South Wales Digital Filter Bank used for the observations with the Mopra Telescope was provided with support from the Australian Research Council. 
This work is financially supported by Grant-in-Aid for Scientific Research (KAKENHI) from Japan Society for the Promotion of Science (JSPS) (Nos.\ 24224005, 23403001, 23006148-01, 22740119, 22540250, 22244014, and 23740149-1), young researcher overseas visits program for vitalizing brain circulation (no.\ R2211) from JSPS, and the Grant-in-Aid for Nagoya University Global COE Program, ``Quest for Fundamental Principles in the Universe: From Particles to the Solar System and the Cosmos'', from the Ministry of Education, Culture, Sports, Science and Technology of Japan (MEXT).
This work is based in part on archival data obtained with the Spitzer Space Telescope, which is operated by the Jet Propulsion Laboratory (JPL), California Institute of Technology under a contract with NASA. Support for this work was provided by an award issued by JPL/Caltech.

\appendix

\section {VELOCITY DISTRIBUTION OF EXPANDING SHELL}\label{sec:expandingshell}

\begin{figure*}
\includegraphics{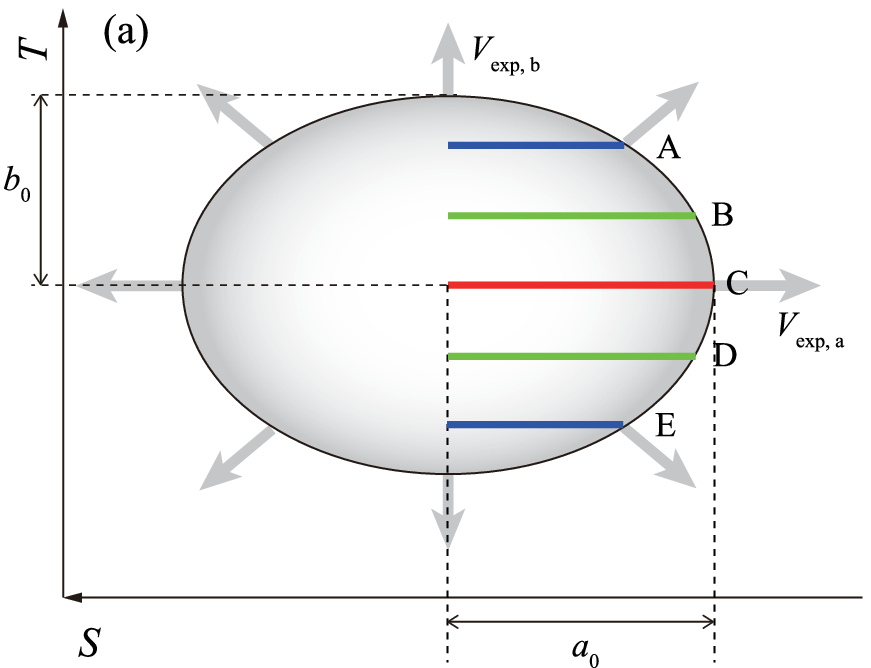}
\includegraphics{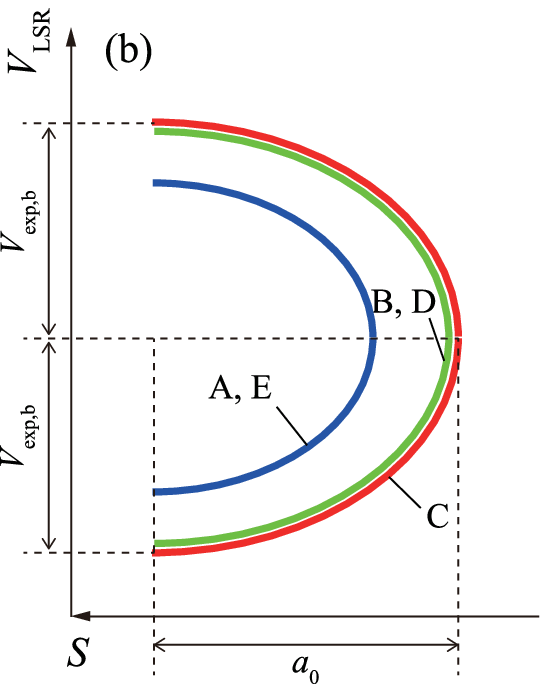}
\caption{ (a) Schematic view of an expanding ellipsoidal shell.
(b) Velocity distributions in position ($S$)-velocity diagram.
\label{expand}}
\end{figure*}

Figure \ref{expand} (a) shows a schematic view of an expanding ellipsoidal shell.
Assuming the {\it arc} cloud and \ion{H}{1} shell are part of an expanding shell, we model the shell to be an ellipsoid with a semimajor axis $a_0$ of 45\,pc, semiminor axis $b_0$ of 32\,pc and a 3rd axis in the depth direction which has the same length as the minor axis (prolate spheroid).
Figure \ref{expand} (b) shows position($S$)-velocity diagrams sliced along the lines of A -- E in Figure \ref{expand} (a).
If the shell is expanding with velocities of $V_{\rm exp,a}:V_{\rm exp,b}=45:32$, an elliptical feature is expected in the position($S$)-velocity diagram as shown in Figure \ref{expand} (b).
The velocity distribution systematically shifts with the projected radius from the center of the shell as shown by red, green and blue colors.
Note that the distributions in Figure \ref{expand} (b) are represented only for the western half.

\section {EVOLUTION OF SNR}\label{sec:snrevolution}

A simple model of the evolution of SNRs is explained by three phases; the free expansion phase in which the shock velocity is constant, the Sedov-Taylor phase in which the shock expands adiabatically, and the radiative phase in which the energy is lost by radiation \citep[e.g.,][]{1997ApJ...490..619S, 2006MNRAS.371.1975Y}.
The shock velocity $V_{\rm s}$ for each phase is written by function of the age of SNRs $t$ as follows;
\begin{equation}
V_{\rm s}(t) = \left\{ \begin{array} {l l l}
		v_{\rm i}																								& (0 \leq t < t_{1})		& \mbox{: Free expansion phase}\\
		v_{\rm i} \left( \frac{t}{t_1} \right)^{-3/5} 												& (t_1 \leq t < t_2)	& \mbox{: Sedov-Taylor phase}\\
		v_{\rm i} \left( \frac{t_2}{t_1} \right)^{-3/5} \left( \frac{t}{t_2} \right)^{-2/3}	& (t_2 \leq t)			& \mbox{: Radiative phase}
  \end{array} \right.
\end{equation}
where $v_{\rm i}$ is the initial shock velocity when the supernova explosion occurs. $t_1$ and $t_2$ are transition time from the free expansion phase and Sedov-Taylor phase, respectively, to the next phases, and are written as $t_1 = 2.1 \times 10^2 E_{51}^{1/3} n_0^{-1/3} v_{\rm i, 4}^{-5/3}$\,yr and $t_2 = 4 \times 10^4 E_{51}^{4/17} n_0^{-9/17}$\,yr \citep{1998ApJ...500..342B}.
$E_{51}$ is an expansion energy in unit of $10^{51}$\,erg, and $n_0$ is an uniform ambient density in unit of cm$^{-3}$. $v_{\rm i, 4}$ is given from $v_{\rm i}=v_{\rm i, 4} \times 10^4$\,km\,s$^{-1}$.
For the radiative phase, we use that the evolution of the shock velocity obeys $\propto t^{-2/3}$ approximately in the epoch of 10$^5$\,yr \citep{1998ApJ...500..342B, 2004A&A...419..419B}.

Radius of an SNR are derived by integrating (B1) along the time, i.e. $R(t)= \int V_{\rm s}(t) dt$.
We show the radii in the free expansion phase $R_{\rm f}$, the Sedov-Taylor phase $R_{\rm s}$, and the radiative phase $R_{\rm f}$ in the follows,
\begin{equation}
R(t)=\left\{
\begin{array}{ll}
R_{\rm f}(t) = v_{\rm i} t & (0 \leq t < t_{1}) \\
R_{\rm s}(t) = \frac{5}{2} v_{\rm i}  \left( t_1^{3/5} t^{2/5} -t_1 \right) + R_{\rm f}(t_1) & (t_1 \leq t < t_2) \\
R_{\rm r}(t) = 3 v_{\rm i} t_1^{3/5} \left( t_2^{1/15} t^{1/3} -t_2^{2/5} \right) + R_{\rm s}(t_2) & (t_2 \leq t)
\end{array} \right.
\end{equation}
Mass swept by the SNR $M_\mathrm{swept}$ is estimated by
\begin{equation}
M_\mathrm{swept} = \frac{4 \pi}{3} R(t)^3 m_{\rm H} n_0.
\end{equation}

\bibliographystyle{apj} 
\bibliography{references}

\clearpage


\clearpage

\clearpage

\clearpage

\clearpage

\clearpage

\clearpage

\clearpage

\clearpage

\clearpage

\clearpage

\clearpage

\clearpage

\clearpage

\clearpage

\clearpage

\clearpage

\clearpage

\clearpage

\clearpage



\begin{table*}
\caption{Observed transitions and telescopes}
\label{prop_tel}
\begin{tabular}{lrrrcc} 
\tableline \tableline
\multicolumn{1}{c}{Line}						& \multicolumn{1}{c}{HPBW}	& \multicolumn{1}{c}{Velocity resolution}	& $T_{\rm rms}$	& \multicolumn{1}{c}{Telescope}	& \multicolumn{1}{c}{Mode\tablenotemark{a}}	\\
								& \multicolumn{1}{c}{(arcsec)}	& \multicolumn{1}{c}{(km\,s$^{-1}$)}			&  \multicolumn{1}{c}{(K)}				& 					& 									\\ \tableline
$^{12}$CO($J$=2--1)	& 100		& 0.33						& 0.2					& NANTEN2	& OTF							\\
$^{13}$CO($J$=2--1)	& 100		& 0.34						& 0.09				& NANTEN2	& PSW							\\ 
$^{12}$CO($J$=1--0)	& 47			& 0.088						& 1.0					& Mopra		& OTF							\\
$^{13}$CO($J$=1--0)	& 47			& 0.092						& 0.4					& Mopra		& OTF							\\ \tableline 
\end{tabular}
\tablenotetext{a}{OTF and PSW represent on-the-fly and the position switching mode, respectively.}
\end{table*}

\begin{table*}
\caption{Observed parameters of $^{12}$CO($J$=2--1) with NANTEN2}
\label{obs_param}
\begin{tabular}{lrrrrrrrrrr} 
\tableline \tableline
Region	& \multicolumn{4}{c}{Position}			&& \multicolumn{4}{c}{Peak properties} 																								\\ \cline{2-5} \cline{7-10}
			&  \multicolumn{1}{c}{$l$} & \multicolumn{1}{c}{$b$} & \multicolumn{1}{c}{$S$} & \multicolumn{1}{c}{$T$} 				&& \multicolumn{1}{c}{$T_{\rm MB}$}		& \multicolumn{1}{c}{$V_{\rm LSR}$}	& \multicolumn{1}{c}{$\Delta V$}			& \multicolumn{1}{c}{$W_{\rm CO}$}		& \multicolumn{1}{c}{$\Delta V_{\rm comp}$} \\
			& \multicolumn{4}{c}{(degree)}			&& \multicolumn{1}{c}{(K)} 						& \multicolumn{1}{c}{(km\,s$^{-1}$)}	& \multicolumn{1}{c}{(km\,s$^{-1}$)}		& \multicolumn{1}{c}{(K\,km\,s$^{-1}$)}	& \multicolumn{1}{c}{(km\,s$^{-1}$)}				\\ \tableline 
Arc		& 283.97	& $-$0.15 & $-$0.35	&  0.02	&& 2.9 						& 26.3 				& 4.8 					& 15.2 					& 2.6								\\ \tableline
J1 		& 284.32	& $-$0.51 &   0.15	& $-$0.03	&& 1.7						& 29.0				& 1.9						& 3.5						& 2.4								\\ 
J2 		& 284.64	& $-$0.73 &   0.54	&   0.00	&& 1.6						& 28.8				& 4.4						& 6.6						& 4.9		 						\\ 
J3 		& 284.94	& $-$0.87 &   0.85	&   0.09	&& 4.5						& 26.2				& 2.4						& 11.7					& 4.0								\\ 
J4 		& 284.60	& $-$0.69 &   0.48	&   0.00	&& 2.5						& 22.5				& 3.7						&  8.7					& 3.4								\\ 
J5 		& 284.83	& $-$0.78 &   0.72	&   0.08	&& 4.1						& 21.3				& 3.5						& 14.7					& 3.7								\\ \tableline
\end{tabular}
\tablecomments{Definition of J1 -- J5 is referred in Figure \ref{finding_chart} and Section \ref{subsubsec:global_distribution}. $T_{\rm MB}$ is the main beam temperature, $V_{\rm LSR}$ central velocity, $\Delta V$ linewidth, $W_{\rm CO}$ total integrated intensity at the peak position of each cloud, and $\Delta V_{\rm comp}$ composite (averaged) linewidth for the entire region of each cloud. $T_{\rm MB}$, $V_{\rm LSR}$ and $\Delta V$ are determined by fitting the spectra with Gaussian. $\Delta V_{\rm comp}$ is obtained by the Gaussian fitting for the averaged line profiles where the integrated intensity is greater than 4$\sigma$.}
\end{table*}

\begin{table*}
\caption{Observed parameters of $^{12}$CO($J$=1--0) with Mopra Telescope}
\label{obs_para12co10}
\begin{tabular}{lrrrrrrrrrr} 
\tableline \tableline
Region	& \multicolumn{4}{c}{Position}			&& \multicolumn{4}{c}{Peak properties} 																								\\ \cline{2-5} \cline{7-10}
		&  \multicolumn{1}{c}{$l$} & \multicolumn{1}{c}{$b$} & \multicolumn{1}{c}{$S$} & \multicolumn{1}{c}{$T$} 				&& \multicolumn{1}{c}{$T_{\rm MB}$}		& \multicolumn{1}{c}{$V_{\rm LSR}$}	& \multicolumn{1}{c}{$\Delta V$}			& \multicolumn{1}{c}{$W_{\rm CO}$}		& \multicolumn{1}{c}{$\Delta V_{\rm comp}$} \\
		& \multicolumn{4}{c}{(degree)}			&& \multicolumn{1}{c}{(K)} 						& \multicolumn{1}{c}{(km\,s$^{-1}$)}	& \multicolumn{1}{c}{(km\,s$^{-1}$)}		& \multicolumn{1}{c}{(K\,km\,s$^{-1}$)}	& \multicolumn{1}{c}{(km\,s$^{-1}$)} 				\\ \tableline 
Arc	& 283.97	& $-$0.15 & $-$0.36	&  0.02	&&  6.4						& 26.4				& 4.1						& 26.4					& 2.5								\\ \tableline
J1 	& 284.45	& $-$0.59 &  0.30	& $-$0.01	&&  4.0						& 28.7				& 2.5						& 10.9					& 2.3								\\
J2 	& 284.62	& $-$0.72 &  0.52	& $-$0.01	&&  3.6						& 28.8				& 3.7						& 13.2					& 4.8								\\
J3 	& 284.94	& $-$0.87 &  0.85	&  0.08	&& 13.7						& 26.2				& 2.2						& 32.7					& 4.1								\\
J4 	& 284.58	& $-$0.70 &  0.47	& $-$0.01	&&  6.7						& 22.9				& 2.6						& 18.2					& 3.6								\\
J5 	& 284.83	& $-$0.79 &  0.72	&  0.08	&&  9.3						& 21.2				& 2.7						& 26.6					& 3.7								\\ \tableline
\end{tabular}
\tablecomments{Meaning of each parameters are the same as in Table 2. $\Delta V_{\rm comp}$ is obtained by the Gaussian fitting for the averaged line profiles where the integrated intensity is greater than 4$\sigma$.}
\end{table*}

\begin{table*}
\caption{Observed parameters of $^{13}$CO($J$=1--0) with Mopra Telescope}
\label{obs_para13co10}
\begin{tabular}{lrrrrrrrrrr} 
\tableline \tableline
Region	& \multicolumn{4}{c}{Position}			&& \multicolumn{4}{c}{Peak properties} 																								\\ \cline{2-5} \cline{7-10}
		&  \multicolumn{1}{c}{$l$} & \multicolumn{1}{c}{$b$} & \multicolumn{1}{c}{$S$} & \multicolumn{1}{c}{$T$} 				&& \multicolumn{1}{c}{$T_{\rm MB}$}		& \multicolumn{1}{c}{$V_{\rm LSR}$}	& \multicolumn{1}{c}{$\Delta V$}			& \multicolumn{1}{c}{$W_{\rm CO}$}		& \multicolumn{1}{c}{$\Delta V_{\rm comp}$} \\
		& \multicolumn{4}{c}{(degree)}			&& \multicolumn{1}{c}{(K)} 						& \multicolumn{1}{c}{(km\,s$^{-1}$)}	& \multicolumn{1}{c}{(km\,s$^{-1}$)}		& \multicolumn{1}{c}{(K\,km\,s$^{-1}$)}	& \multicolumn{1}{c}{(km\,s$^{-1}$)} 				\\ \tableline 
Arc	& 283.97	& $-$0.15 & $-$0.35	&  0.02	&& 2.4						& 26.1				& 2.1						& 5.4						& 1.9								\\ \tableline
J1 	& 284.45	& $-$0.57 &  0.29	&  0.00	&& 0.8						& 28.3 				& 1.7						& 1.8						& 2.1								\\
J2 	& 284.62	& $-$0.73 &  0.52	& $-$0.01	&& 0.3						& 29.1				& 2.4						& 0.8						& 3.9								\\
J3 	& 284.94	& $-$0.86 &  0.85	&  0.09	&& 3.3						& 26.2				& 1.7						& 6.2						& 2.4								\\
J4 	& 284.58	& $-$0.69 &  0.46	& $-$0.01	&& 1.7						& 23.0				& 1.4						& 2.5						& 2.1								\\
J5 	& 284.84	& $-$0.79 &  0.72	&  0.08	&& 2.5						& 21.2				& 2.1						& 5.7						& 2.7								\\ \tableline 
\end{tabular}
\tablecomments{ Parameters in this table were obtained by binding up two channels to decrease the noise. Meaning of each parameters are the same as in Table 2. $\Delta V_{\rm comp}$ is obtained by the Gaussian fitting for the averaged line profiles where the integrated intensity is greater than 3$\sigma$.}
\end{table*}

\begin{table}
\caption{Physical properties of molecular clouds}
\label{phys_param}
\begin{tabular}{lrrr} 
\tableline \tableline
\multicolumn{1}{c}{Region}	& \multicolumn{1}{c}{peak $N({\rm H_2})$\tablenotemark{a}}				& \multicolumn{1}{c}{$r$\tablenotemark{b}}	& \multicolumn{1}{c}{$M$\tablenotemark{c}} 	\\
			& \multicolumn{1}{c}{(10$^{21}$\,cm$^{-2}$)}	& \multicolumn{1}{c}{(pc)}							& \multicolumn{1}{c}{($10^4\,M_\sun$)}							\\ \tableline
Arc 		& 2.7								& 18.6							& 2.5												\\ 
WJ		& 1.4								& 7.2								& 0.3												\\
J1		& 1.7								& 9.9								& 0.6												\\
J2 		& 2.7								& 11.6							& 1.0												\\
J3 		& 4.8								& 15.9							& 2.1												\\
J4		& 2.9								&  8.2							& 0.5												\\
J5		& 4.8								& 11.7							& 1.4												\\ \tableline
\end{tabular}
\tablenotetext{a}{H$_2$ column density at the integrated intensity peak.}
\tablenotetext{b}{Radii of J4 and J5 are derived by the integrated intensity maps of $^{12}$CO($J$=2--1) above 4$\sigma$.}
\tablenotetext{c}{Mass of J4 and J5 are derived by converting the integrated intensities of $^{12}$CO($J$=2--1) into those of $^{12}$CO($J$=1--0) with mean integrated intensity ratios of 0.43 for J4 and of 0.47 for J5 in the regions detected in $^{12}$CO($J$=2--1) above 4$\sigma$.}
\end{table}

\begin{table*}
\caption{Observed parameters used in LVG analysis and results of LVG analysis.}
\label{tlvgprop}
\begin{tabular}{lrrrrcrrrrrr} 
\tableline \tableline
Region	& \multicolumn{4}{c}{Position}			&& \multicolumn{3}{c}{Averaged intensities\tablenotemark{a}}	&						& \multicolumn{2}{c}{LVG results} 						\\ \cline{2-5}  \cline{7-9} \cline{11-12}
			& \multicolumn{1}{c}{$l$}			& \multicolumn{1}{c}{$b$}	& \multicolumn{1}{c}{$S$}	& \multicolumn{1}{c}{$T$}	&& \multicolumn{1}{c}{$T_{\rm obs1}$}	& \multicolumn{1}{c}{$T_{\rm obs2}$}	& \multicolumn{1}{c}{$T_{\rm obs3}$}		& \multicolumn{1}{c}{$V_{\rm G}$}	& \multicolumn{1}{c}{$n({\rm H}_2)$}							&$T_{\rm kin}$	\\ 
			&\multicolumn{4}{c}{(degree)}			&& \multicolumn{3}{c}{(K)}													& \multicolumn{1}{c}{(km\,s$^{-1}$)} 	& \multicolumn{1}{c}{(cm$^{-3}$)}								&\multicolumn{1}{c}{(K)}					\\ \tableline
p			& 283.97	& $-$0.15	& $-$0.35	&  0.02	&& 2.63 				& 1.73 					& 0.91 						& 26.0				& $1.4_{-0.1}^{+0.1} \times 10^3$	& $9_{-2}^{+2}$	\\
q   		& 284.27	& $-$0.43	&  0.06	&  0.00	&& 1.38 				& 0.48 					& 0.11 						& 29.2 				& $\le 5.0 \times 10^2$				& $\le 6 $			\\
r   		& 284.32	& $-$0.53	&  0.16	& $-$0.04	&& 1.22 				& 0.35 					& 0.12 						& 29.1 				& $5.6_{-0.1}^{+0.0} \times 10^2$	& $9_{-1}^{+1}$	\\
s   		& 284.45	& $-$0.58	&  0.29	& $-$0.01	&& 0.89 				& 0.44 					& 0.10 						& 28.6 				& $\le 5.9 \times 10^2$				& $\le 6$			\\
t-1		& 284.62	& $-$0.73	&  0.51	& $-$0.01	&& 1.06 				& 0.25 					& 0.11 						& 24.2 				& $5.9_{-0.2}^{+0.1} \times 10^2$	& $13_{-2}^{+2}$	\\		
t-2		&        		&			&			&			&& 1.09 				& 0.22 					& 0.10 						& 28.8 				& $5.6_{-0.2}^{+0.1} \times 10^2$	& $15_{-3}^{+3}$	\\
u-1		& 284.77	& $-$0.81	&  0.68	&  0.02	&& 1.16 				& 0.20 					& 0.11 						& 22.2 				& $5.9_{-0.3}^{+0.2} \times 10^2$	& $20_{-5}^{+5}$	\\
u-2		&        		&			&			&			&& 1.90 				& 0.46 					& 0.19 						& 26.2 				& $5.7_{-0.1}^{+0.2} \times 10^2$	& $12_{-2}^{+2}$	\\
v			& 284.75	& $-$0.98	&  0.78	& $-$0.13	&& 1.36 				& 0.27 					& 0.16 						& 26.4 				& $7.2_{-0.3}^{+0.6} \times 10^2$	& $21_{-5}^{+7}$	\\
w			& 284.93	& $-$0.87	&  0.85	&  0.08	&& 4.10 				& 1.55 					& 1.08 						& 26.5 				& $1.3_{-0.2}^{+0.2} \times 10^3$	& $16_{-4}^{+6}$	\\ \tableline
\end{tabular}
\tablenotetext{a}{$T_{\rm obs1}$, $T_{\rm obs2}$ and $T_{\rm obs3}$ represent the observed intensities in $^{12}$CO($J$=2--1), $^{13}$CO($J$=1--0) and $^{13}$CO($J$=2--1), respectively, averaged in 1 km\,s$^{-1}$ width around the barycentric velocities $V_{\rm G}$.}
\end{table*}

\begin{table}
\caption{AKARI  sources}
\label{akari}
\begin{tabular}{llrc} 
\tableline
\tableline
\multicolumn{1}{c}{Catalog} 	& \multicolumn{1}{c}{ID} 							& \multicolumn{1}{c}{Band} 			& Flux density 			\\
				&								& \multicolumn{1}{c}{({\micron})}	& (Jy) 						\\ \hline
IRC PSC	& 1026103-583340	& 9				& $0.313 \pm 0.019$ 	\\
FIS BSC	& 1026099-583345	& 160			& $17.8 \pm 7.6$ 		\\ \tableline
\end{tabular}
\end{table}

\end{document}